\documentclass[doc]{apa7}
\usepackage[T1]{fontenc}
\sffamily

\usepackage{lipsum}
\usepackage{ragged2e}
\usepackage[american]{babel}

\usepackage{csquotes}
\usepackage[section]{placeins}

\usepackage[style=apa,sortcites=true,sorting=nyt,backend=biber]{biblatex}
\DeclareLanguageMapping{american}{american-apa}
\title{Network Models of Expertise \\in the Complex Task of Operating Particle Accelerators}
\shorttitle{Expertise in Particle Accelerators}

\authorsnames[{1,2},2,2,2]{Roussel Rahman, Jane Shtalenkova, \\Aashwin Ananda Mishra, Wan-Lin Hu}
\authorsaffiliations{{Stanford University, Stanford, CA},{SLAC National Accelerator Laboratory, Menlo Park, CA}}

\leftheader{Rahman}

\abstract{
We implement a network-based approach to study expertise in a complex real-world task: operating particle accelerators. Most real-world tasks we learn and perform (e.g., driving cars, operating complex machines and devices, solving mathematical problems) are difficult to learn because they are complex, and the best strategies are difficult to find from many possibilities. However, how we learn such complex tasks remains a partially solved mystery, as we cannot explain how the strategies evolve with practice due to the difficulties of collecting and modeling complex behavioral data. As complex tasks are generally networks of many elementary subtasks, we model complex task performance as networks or graphs of subtasks and investigate how the networks change with expertise. We develop the networks by processing the text in a large archive of operator logs from 14 years of operations using natural language processing and machine learning. The network changes are examined using a set of measures at four levels of granularity – individual subtasks, interconnections among subtasks, groups of subtasks, and the whole complex task. We find that the operators consistently change with expertise at the subtask, the interconnection, and the whole-task levels, but they show remarkable similarity in how subtasks are grouped. These results indicate that the operators of all stages of expertise adopt a common divide-and-conquer approach by breaking the complex task into parts of manageable complexity, but they differ in the frequency and structure of nested subtasks. Operational logs are common data sources from real-world work environments where people collaborate with hardware and software environments to execute complex tasks, and the network models investigated in this study can be expanded to accommodate multi-modal data. Therefore, our network-based approach provides a practical way to investigate expertise in the real world.

}


\authornote{
   \addORCIDlink{Roussel Rahman}{0000-0003-2009-4246}

  Correspondence concerning this article should be addressed to Roussel Rahman, Department of Photon Science, SLAC National Accelerator Laboratory, Menlo Park, CA.  E-mail: \href{mailto: roussel.rahman@gmail.com}{roussel.rahman@gmail.com}}

\begin{document}

\maketitle
\newpage
The manner in which humans learn new tasks is of substantial interest in Psychology, as well as other fields such as Neuroscience \parencite{bassett2017network}, Machine Learning \parencite{shin2021human, shin2023superhuman}, etc. Many pioneering psychologists, including Jean Piaget, Lev Vygotsky, Allen Newell, Herbert Simon, and John Anderson, have enriched our understanding of human learning. The reason for the interest is simple: learning is a fundamental process for all humans and our understanding of learning may help us find ways to accelerate learning and maximize our potential. 

Despite this progress, the manner by which humans learn complex, real-world tasks remains less explored due to the inherent difficulties in conducting experiments in real-world work settings. Most theories of learning have been developed by studying learning relatively simpler tasks (e.g., memory or reaction time tasks, mental arithmetic tasks) that represent tests of fundamental skills \parencite{newell1981mechanisms, anderson1982acquisition, rickard1997bending, delaney1998strategy, logan2002instance, donner2015piecewise}. In contrast, real-world tasks are almost always complex. If we consider the tasks of commuting to work, writing a program, or solving a mathematical problem, there are essentially infinite possible paths and sets of actions to choose from for each task. The space of possible strategies becomes too large for both humans and machines to search for optimal strategies exhaustively. The manner in which humans tackle the complexity of finding appropriate strategies to master such tasks remains a partially solved problem.

To study how our strategies evolve with learning and practice, we need new mathematical approaches to model complex behavioral data. The changes in strategy may manifest as changes in subtask selection, in interconnections between subtasks, or in how subtasks are grouped into communities. As complex tasks can be represented as networks or graphs of many subtasks, Network Science – which is built upon Graph theory – provides exceptionally suitable tools to represent and investigate the learning of complex tasks. In this work, we model complex task performance as networks of subtasks and investigate the changes in strategy with expertise through the changes in the structure and the connectivity of these networks. Thereafter, we examine performance changes with expertise at four different network levels – (1) individual subtasks, (2) the interconnections between subtasks, (3) the groups of closely connected subtasks, and (4) the whole complex network of subtasks.

We implement and test the network-based approach to investigate the dynamics of expertise in an extremely complex real-world task: tuning the Linac Coherent Light Source (LCLS) at SLAC National Accelerator Laboratory. This linear particle accelerator serves as a Free-Electron Laser (FEL) light source of extremely bright X-rays for researchers from all over the world, who use these X-rays to illuminate samples and capture high-resolution images and videos at the particle scale (e.g., atomic motions, chemical reactions, and molecular structures of materials) and even to ``create matter from the vacuum'' \parencite[p. 33]{pellegrini2003x}. However, the researchers do not interface directly with the source accelerator. Instead, a team of highly trained and skilled operators configure and tune the LCLS to meet the desired specifications of an ever-changing schedule of scientific experiments. 

The manipulation of the LCLS charged particle beam to generate the exact energy, bandwidth, and brightness of X-rays required for each experiment is called FEL tuning. We examine the effects of differing levels of expertise in FEL tuning using the electronic logs (or elogs) created by operators during the 14 years since the start of LCLS operations at SLAC. These elogs are processed using Machine Learning and Natural Language Processing (NLP) techniques to extract information on FEL Tuning and its subtasks. In turn, we use this information to create network models of FEL performance for operators of varying stages of expertise and investigate the changes using a host of network science based tools and algorithms.

In the next section, we briefly review previous works on complex task solving and learning. In Section \ref{sec:new_promising_approaches}, we discuss modeling complex task performance as networks and investigating the changes with expertise by the changes in the networks. Our research questions are specified in Section \ref{sec:research_questions} before describing our experimental task – FEL tuning -- in Section \ref{sec:fel_tuning}. The methods we used to construct the network models and measure the changes in the networks are described in Section \ref{sec:methods}. The results, discussion, and conclusions are presented respectively in Sections \ref{sec:results}, \ref{sec:discussions}, and \ref{sec:conclusions}.

\section{Solving and Learning Complex Tasks}
Generally, complex tasks consist of many relatively simpler subtasks organized in hierarchical networks \parencite{simon1962architecture}. In these networks, the subtasks at higher levels serve as the goals for those at lower-levels. For example, taking a turn while driving is a high-level task that serves as the goal for several lower-level subtasks, such as checking traffic signals, pressing the brake, and rotating the steering wheel. The task of making a turn is itself a subtask of the even higher-level task of driving a car. Due to the interconnections among the subtasks, there are many possible paths to reach the goals at the highest levels from the lowest-level actions, too many to find the optimal choice through an exhaustive search of the parameter space. In this context, the key questions involve investigating the manner by which humans learn and organize these subtasks to solve complex problems, as well as the details of how subtask selection and organization strategy evolve with experience and expertise. 


\subsection{Information-based Explanations}

Information-based explanations focus on the information content used in solving problems and the representations of the information in human memory. The most prominent theory is the chunking theory of information processing, which posits that humans selectively attend to and store information as chunks. These chunks are compact/condensed versions of the original information and enable the storage of more information within limited memory. Chunking has been used to explain expertise in many complex tasks, for instance, Chess \parencite{chase1973perception}, air-traffic control \parencite{lee2001does, vanmeeuwen2014identification}, Cyber Security \parencite{cranford2021towards, aggarwal2016cyber}, and even video gameplay \parencite{thompson19plosOne}.

The chunking theory is a foundation for the Adaptive Control of Thought-Rationale (ACT-R), the most complete cognitive architecture to date \parencite{anderson1997act, anderson2013implications, sun2008introduction}. 
The ACT-R considers the information content (as chunks) and the decision processes (as rules) based on the information. With the acquisition of new chunks, a learner may also learn new rules, and these changes drive improvements with practice. The instance-based learning (IBL) theory \parencite[]{gonzalez2003instance}, developed based on the ACT-R architecture, extends the concept of chunks to instances with specific attributes (e.g., state of environment, action, and utility). On each trial, the instances (or the chunks for the ACT-R) from memory are retrieved according to their activation values, using the current state as the retrieval cue. The activation values are used to estimate the probability distributions over all retrieved instances and calculate the expected utility of the possible actions \parencite{gonzalez2003instance, gonzalez2011instance, gonzalez2013boundaries}. Then, the action with the maximum expected utility is chosen for the current trial. Both the ACT-R and the IBL theory have been successfully applied to explain learning in a wide range of simple and complex tasks, including cybersecurity, airport security, team performance in collaborative games, and more \parencite[For reviews, please see][]{gonzalez2011instance, ritter2019act}.

Another information-focused perspective on the representation and learning of complex tasks is the mental model theory, which states that humans create mental models that are simplified representations of actual systems that are too complex to retain in full. With experience, the individuals’ mental models improve, which results in performance improvements. Mental models were first proposed for individuals before being extended to explain team performance. Many experimental studies demonstrate that team members need shared mental models for synchronizing actions and maximizing the achievement of shared goals \parencite[e.g.,][]{mathieu2000influence, cannon1993shared, andrews2023role}. Recent applications include studying alignment between human and AI teammates and contributing to developing explainable AI models to improve synchronization between and performance of human-AI teams \parencite[for a review, please see][]{andrews2023role}.


Despite the success, a common roadblock for real-world application is that most of the investigations to date have been in laboratory experiments using simple tasks. In these tasks, the researchers know \textit{a priori} the chunks of information that learners can use, and this knowledge can be used to develop and investigate information-processing models of human behavior. For more complex tasks in the real world, knowing what information is relevant to performance and how it changes with expertise requires considerable domain knowledge. This requirement of domain knowledge is also true for eliciting mental models based on user interviews, surveys, and think-aloud protocols \parencite[]{andrews2023role}. Overall, the methods of analysis have been varied and often domain-specific. General methods to (1) identify task-relevant information and (2) quantitatively study their changes can help in extending these theories of learning and generalize for more complex real-world tasks.

\subsection{Process-based Explanations}

Process-based explanations focus on the manner in which humans solve problems within the bounds or the limits of their computational capabilities. In simple laboratory tasks, the learners and the researchers both have access to complete information to identify the best methods. 
However, in complex real-world tasks, finding the appropriate methods from many possible alternatives is a complex problem nested within the tasks.

In solving complex problems, humans “satisfice” (satisfy and suffice) using heuristics – that is, they approximate good enough solutions that satisfy a set of conditions and suffice for their current goals – instead of trying to optimize \parencite{newell1958elements, simon1971human, simon1997models, gigerenzer1996reasoning, gigerenzer2008heuristics, gigerenzer2020bounded}. The goal of optimizing provides an all-or-none approach to search for solutions that cannot be implemented in complex problems \parencite{simon1975optimal}. In contrast, heuristic search consists of iteratively executing a set of steps that can be represented as a production system or, more simply, as a set of if-else statements. For example, satisficing as a heuristic consists of (1) setting an aspiration level that will suffice for our current needs, (2a) a criterion to stop search if we find a satisfactory solution, (2b) else stop search if we fail to find one within a number of trials, and (3) adjust our aspiration levels accordingly \parencite{ gigerenzer2008heuristics, gigerenzer2020bounded}. With optimality beyond reach in complex problems, heuristics provide frugal and efficient alternatives in terms of computational time and effort, and they often outperform more sophisticated algorithms in complex and uncertain environments; for example, in investing in the stock market, driving a car, or flying a plane; please see \textcite{gigerenzer2008heuristics} for an elaborate review. 

Satisficing using heuristics is a bounded, rational approach that is in sharp contrast to the rationality principle of maximizing expected utility – which defines rational behavior as optimal behavior by choosing the option with the maximum expected utility \parencite{simon1976substantive, Simon1992TheGO, gigerenzer2020bounded, gigerenzer2020explain}. The principle of utility maximization is assumed in almost all of Economics and a large part of Psychology, but it may provide infeasible solutions for complex tasks.
\textcite{Simon1992TheGO} provide an example using the game theory for Chess expertise. According to game theory, the optimal strategy for playing chess is first to draw the tree of all possible states in chess games, then simply follow the path that provides the maximum expected utility (i.e., a win). However, as the number of possible states in Chess is extremely large ($\mathcal{O}(10^{120})$ as estimated by \cite{shannon1948mathematical}), this strategy cannot be implemented within our computational limits. Therefore, it is not a bounded rational solution.

Other similar method-based explanations of human behavior include the concepts of schemas and routines \parencite[For reviews, please see][]{nathan2014foundations}. Apart from some nuanced differences, the general idea is that, over the course of learning, humans search for and find improved methods that map their actions to goals. This view has been captured in the theory of learning by doing \parencite{anzai1979theory} – a general theory of human learning in well-structured complex problems with specific goals and sets of alternatives to achieve the goals. According to the theory, humans selectively search for information with each trial, develop new subgoals to reach using the information, and iteratively improve methods to achieve the subgoals. They demonstrated the theory's accuracy by predicting and modeling the method changes of a new learner in the Tower of Hanoi task (a well-structured complex problem). 

An advantage the researchers had in these studies is the foreknowledge of all possible actions and the typical subgoals to achieve in the task based on previous research. For most real-world tasks, such information is not available. Moreover, real-world tasks are often ill-structured as they may not have specific sets of alternative actions or sometimes even specific goals to reach, such as in engineering or architectural design \parencite{simon1998we}. An additional step in learning such tasks is generating a set of alternatives and finding the (sub)goals; as the goals change throughout learning, the appropriate methods will also change \parencite{anzai1979theory, simon1998we, rahman2022dynamics, gray2017pdls}. A general theory of complex task learning needs to explain the changes of methods over the course of learning, as the theory of learning by doing does for the class of well-structured problems.

\section{Towards New Approaches to Study Complex Task Learning}
\label{sec:new_promising_approaches}
\subsection{The Need to Capture the Method Changes While Learning}

Understanding the method changes can help in finding ways to accelerate the transitions through different methods to reach an expert state. It has been observed many times that humans go through a sequence of improved methods as they learn complex tasks \parencite[for examples, please see][] {siegler1987perils, siegler1989children, donner2015piecewise, towne2016understanding, gray2017pdls, rahman_gray2020topics, rahman2021precisemeasures}. On the other hand, humans have also been observed to remain stuck in performance plateaus due to using suboptimal methods even after years of experience in a wide range of tasks, such as typing on keyboards, writing programs, using productivity or engineering software, and playing video games \parencite[for reviews, please see][]{gray2017pdls, gray2017plateaus}. Without considering the complexity of finding and implementing optimal methods, we may dismiss such behavior as irrational. However, considering the behavior as boundedly rational when optimality is beyond reach, we see that a general way to aid humans in learning complex tasks is to help them overcome the complexity of finding appropriate methods \parencite{shin2023superhuman, gaessler2023training}.

Another key point is that a large body of experimental evidence shows that the principles derived from studying simple tasks do not generalize to complex tasks \parencite{wulf2002principles}. The differences often lie in the effects of different training and practice protocols on learning rate. For example, some protocols tend to have opposite effects on simple vs complex task learning, whereas some factors are important only for learning complex tasks. For an elaborate review, please see \textcite{wulf2002principles}. \textcite{lee2001does} directly examined whether the theories developed by studying simple skills can scale up for complex skills by means of relating the high-level changes in performing in an Air-Traffic Control task (e.g., overall accuracy, goals, changes in attention) with the changes at lower levels of performance (e.g., keystrokes, eye-movement). The authors found that the changes at the lowest levels closely reflect those at the highest levels.

An emergent question is whether performance at all levels of a complex task shall exhibit similar changes. We believe it is unlikely since some parts of the whole task may be irrelevant to others. \textcite{gray2017pdls, rahman2021precisemeasures, towne2016understanding} all mention the difficulty of identifying the right scope to observe and measure changes with learning and promote modeling and investigating complex task performance using several measures at different levels to study the changes of methods. Complex tasks have multiple performance levels, and method changes may occur at one or many of these levels. Therefore, we need to model and examine performance and their changes at different levels. We also need to accommodate the possibility of incomplete and sparse information, especially when using naturally occurring data to develop our models. As we discuss and show in this work, network science provides exceptionally suitable tools for these purposes.

\subsection{Performance and Expertise in Complex Tasks through Network Models}
Network science provides a quantitative framework to represent complex cognitive systems and model their changes across multiple levels and timescales \parencite{barabasi2012network, barabasi2013network, borsboom2021network, siew2019cognitive}. Network representations and analyses have been helpful in a wide range of psychological studies, such as semantic and lexical retrieval, creative processes, mental navigation, and more \parencite[for elaborate reviews, please see] []{siew2019cognitive, kenett2020cognitive}. For example, \textcite{griffiths2007google} demonstrated that people’s responses in a semantic retrieval task could be better predicted by PageRank of words – a measure of the importance of the nodes in an interconnected network – than using word frequency or association strengths individually. As for mental navigation and search, it has been found that people navigate and search in their internal cognitive representations of problems in similar ways they navigate in physical spaces \parencite[pp. 159-161]{siew2019cognitive}. Network models have also been helpful in personality research, analyzing voter sentiments, and explaining developmental changes and mental disorders \parencite{goedschalk2018network}.

The key focus of network models is to retain information about the interconnections between the variables or parameters of interest. Therefore, network models provide a direct way to represent complex tasks as networks or graphs of many subtasks and analyze the changes with expertise (or, more generally, across experimental conditions) at different levels of granularity. A network model or a graph consists of a set of nodes or vertices connected by a set of edges $(G: V, E)$. The nodes represent the variables of interest, and the edges represent the relationships between pairs of nodes. In our case, the network nodes represent subtasks, and the relationships between subtasks can be represented as edge attributes. A diverse set of edge attributes can be integrated into one or few networks or kept separate in different networks. The resulting networks can be analyzed by an established set of measures and methods at different levels of aggregation and at different timescales. 

Moreover, complex tasks have been represented in several ways in the past that can be transformed into networks, such as (1) as mazes or grids \parencite{simon1962architecture, simon1971human} and (2) as decision trees \parencite{simon1976strategy_shift}. In both cases, the solutions are represented by paths from the starting point to a desired endpoint. Network models provide a general representation of these approaches and allow the findings to be related to many experimental studies of human problem-solving. A more recent stream of support for graph-based approaches comes from the success of graph-based machine learning and graph neural networks, which have been successfully used to model complex relationships in various domains \parencite{hu2020open}. A reason is that graph-based approaches exploit the network structure for improved performance \cite{hu2020open, zhang2018link}. This is in addition to the inherent features of the nodes used by classical models. The strength of the graph based formalism lies in its ability to incorporate the relationships between objects, in addition to the properties of these objects in isolation.
The current best methods combine the multiple structural features (e.g., node embeddings, subgraphs) to outperform traditional ML methods for node classification and link prediction tasks \parencite{hu2020open}. These findings highlight the promise of graph-based approaches in modeling complex task performance and learning.

We test our network-based approach by modeling performance and expertise in FEL tuning, a complex task that provides a suitable challenge for the methods. In the next section, we describe the specific research questions and hypotheses we seek to address.




\section{Research Questions}
\label{sec:research_questions}

The main exploratory question we address in this work is: How do the strategies for the complex task of tuning the X-ray brightness of the accelerator change with expertise? Changes in strategy with expertise may affect tuning performance, and we can explore these changes at different network levels, from subtasks to the whole problem. To model and investigate the changes at different levels of the complex task, we divide our focal question into two parts:

\begin{enumerate}
    \item What are the subtasks that make up the complex task, and how are they organized?
    \item What changes with expertise at the different levels of the complex task?
    
\end{enumerate}

We will specify these questions for the FEL tuning in the next section. The following are a few hypotheses regarding the answers to these research questions.

\begin{enumerate}
    \item The operators will divide the complex task of X-ray tuning into parts of manageable complexity. 

If true, we will find a strong community structure among the tuning subtasks. 

    \item There will be measurable changes with expertise at different network levels.

If true, we will see our metrics at different network levels change with operator expertise. We expect differences in how subtasks are interconnected and organized into communities and/or changes in the structure of whole networks with expertise.

  


    \end{enumerate}

\section{The Complex Task for Operating Particle Accelerators}
\label{sec:fel_tuning}

In this work, we offer a methodology to passively study complex tasks: we process operational log data created by operators at SLAC National Laboratory as they monitored and configured a particle accelerator. The particular accelerator we studied was the Linear Coherent Light Source (LCLS), a free electron laser (FEL) light source that produces very bright and powerful X-rays at ultra-fast pulse rates. These X-rays enable researchers to look at and take high-resolution, high-frequency snapshots of matter on molecular and even atomic scales. The LCLS facilitates thousands of experiments each year that aid our understanding of phenomena such as photosynthesis~\parencite{photosystem} and molecular interactions for drug discovery~\parencite{drug_discover}. A primary task for operators in the control room is to change the particle accelerator parameters to the specifications required by each experiment using the accelerator control system and then incrementally improve the X-ray laser beam brightness in a process called FEL tuning. 

Operators change the parameters of the machine to deliver variable configurations of the X-ray laser beam within a human-in-the-loop complex control system: a collaboration between human operators, computational algorithms, automated controls, and system experts who work in tandem to operate, troubleshoot, and repair accelerator components to ensure safe and efficient beam delivery. Operators use numerous computer programs and interfaces to collect information and control accelerator states and processes. he operators undergo on-the-job accelerator physics and operations training and learn to recognize the signatures of and known solutions to common problems over time. The training program involves a series of Accelerator Operator training stages, including a year of foundational knowledge and practical skill acquisition until the operator can work with minimal supervision in the control room. 

 The expertise that the operators need to connect domain knowledge and control room actions into solutions for more complex accelerator problems, is developed over time. Three common features of more complex control room problems are that they are (1) extremely complex with very large parameters spaces \parencite{duris2020bayesian, mishra2021uncertainty, gupta2021improving, edelen2018opportunities}, (2) do not have deterministic solutions, and (3) no two problems are exactly alike. FEL tuning is one such complex problem, the goal of which is to set three X-ray beam parameters: (1) Photon energy, (2) Pulse length, and (3) Pulse intensity or beam brightness. The first two parameters are set to specific values requested by the researchers, which are relatively straightforward for the trained operators. The third parameter, beam brightness, can be incrementally increased through the application of a sequence of tuning tweaks - slight modifications of accelerator control parameters to increase the production of X-ray photons from the source electron beam. The parameters adjusted are located geographically along the beamline and impact the beam in different ways. The beam brightness is a moving optimum that operators need to ``tune up'' after changing configurations while also factoring for drift in machine components over time. 

Operators have identified twenty-seven different parameters commonly used for tuning X-ray beam brightness, which we refer to as the tuning parameters from here onward (For a list, please see Appendix \ref{app:tuning_parameters}). Though the parameters that can be adjusted are known, the choice and sequence of tasks is a matter of operator strategy.

To illustrate the task’s complexity, we use two simplified views of methods to estimate the number of possible methods. Considering the methods as ordered lists of the parameters to use, there are $27!= 1.09\times10^{28}$ possible permutations. Considering the methods as subsets of highly connected parameters, there are $\approx 5.45 \times 10^{20}$ possible ways (corresponding to the Bell number of 27) in which a set of 27 elements can be partitioned into subsets or communities.

Evidently, it is not possible for the operators to evaluate the expected utilities of all possible methods. As predicted by the principles of bounded rationality, the operators use a set of heuristics to guide their performance. For example, the goal of maximizing the beam brightness – which in its original form is an optimization problem – is transformed into a satisficing problem by setting an acceptable threshold to achieve based on historical data, current configuration, and time allocated for tuning. A second example is that operators perform trial-and-error search based on feedback -- a form of heuristic search \parencite{simon1975optimal} -- as a common strategy for some of the tuning parameters. Finally, there is a strong reliance on storing and using past solutions that worked for similar problems, reflecting the mirror-the-successful heuristic \parencite{gigerenzer2008heuristics}.

Detailed data on operations is archived in many forms, including digital training articles used to onboard new operators and as a reference in the control room. Another crucial source of information is the operators’ own electronic logs (elogs), that are composed of timestamped text entries cataloging events as they unfold in the control room. As the accelerator operations are too complex to keep in memory, these elogs can be considered as chunks of information the operators deem important enough to store for future use. Therefore, these elogs provide a rare opportunity to investigate the chunks of information used by humans in performing an extremely complex task. In this work, we use the training articles as a reference along with the data in these elogs to model FEL tuning strategy as networks of subtasks and investigate the changes in these networks with expertise.

\section{Methods}
\label{sec:methods}


\subsection{Participants}

\begin{figure}[!ht]
    \caption{Histograms showing (a) the number of operators in each stage of experience and (b) the corresponding number of elog entries. The numbers are shown in half-year increments up to Year 10, after which all operators are binned into one group. In our year-wise investigations, we include periods with at least 50 entries (the red dashed line in Figure b), which is the case up to Year 7.}
    \includegraphics[width=1.0\textwidth]{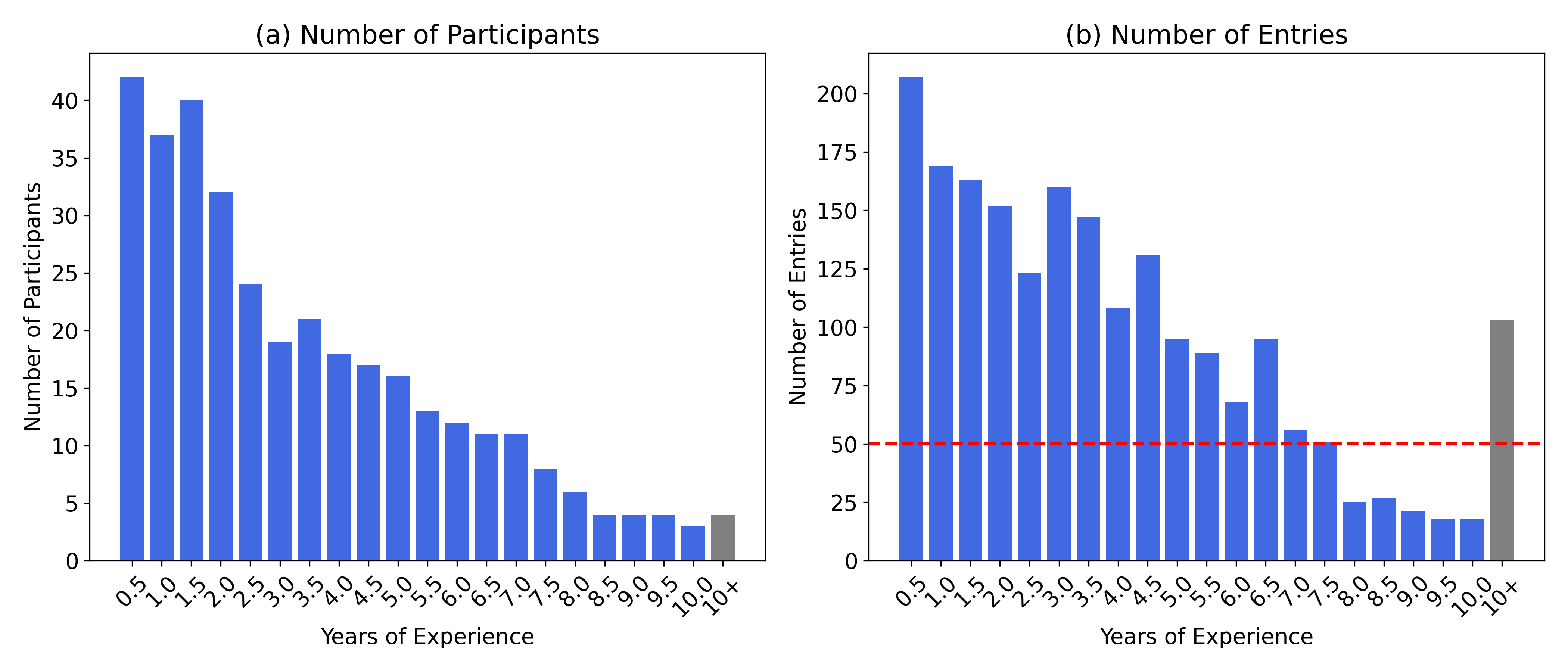}
    \label{fig:participants_barplots}
\end{figure}

Our dataset includes entries from operators along a continuum of training with varying amounts of experience. For each elog entry, we calculated the experience of the operator who authored it by the time elapsed since the operator wrote their first elog entry. In this work, we examine the network changes in two ways: (i) by years of experience (Figure \ref{fig:participants_barplots}) and (ii) by dividing all operators and their entries into three groups of expertise (Table \ref{tab:three_groups}). 

Figure \ref{fig:participants_barplots} shows the number of operators for each stage of experience (in half-year increments) and the number of elog entries they created. The number of operators and the entries both can be observed to decrease with expertise. To have a reasonable amount of data at each stage, we use the criterion of a minimum of 50 entries (the red dashed line in Figure \ref{fig:participants_barplots}b) in a period to be included. Therefore, we include all stages up to Year 7 that satisfy this inclusion criterion.

Table \ref{tab:three_groups} shows the three expertise groups containing all operators and their entries: the Novice group with less than 1 year of experience, the Intermediate group with between 1 and 4 years of experience, and the Expert group with more than 4 years of experience.  These bins are selected to align with existing operator training stages. Note that the number of samples varies considerably across groups. To ensure that the unequal bin sizes do not mislead us, the group-wise investigations are preceded by the examinations of the more continuous changes with years of experience.

\begin{table}[h!]
  \begin{threeparttable}
    \caption{Three groups of operators by stage of expertise}
    \label{tab:three_groups}
    \begin{tabular}{@{}lrrr@{}}         \toprule
    Group                           & Years of Experience & Number of Operators & Number of Entries\\ \midrule
    (1) Novice operators            &    $\leq 1$         & 54                  & 390     \\
    (2) Intermediate operators      &    $1-4$            & 54                  & 876     \\
    (3) Expert operators            &    $\geq 4$         & 26                  & 822     \\ \midrule
    All operators                   &    $0 - 14$         & 67                  & 2088     \\
    \end{tabular}


  \end{threeparttable}
\end{table}

\subsection{The Elog Dataset}
In this work, we use data from two sources – (1) the large database of text elog entries on operations and (2) a detailed operations training article about FEL tuning. 

The elog dataset contains all elog entries from the 14-year period between 2009 and 2022. There are about 350 thousand total entries. Each entry consists of a title and a main text, along with author information, the times and the dates of making the entries, occasional references to preceding/succeeding entries on the same topic, and optional tags for related areas of the accelerators. In our analysis, we combine the title and the main text into a single piece of text. 

For our analysis, we need to subset the entries related to our experimental task of FEL tuning. We also need to identify the parameters discussed in tuning. To do so, we referenced the FEL tuning training article written by expert operators. We determine the similarity between the elog text entries and the tuning article (a) to subset entries related to tuning and also (b) to identify the tuning parameters discussed in each entry in the subset. In the next section, we describe the specific steps we used to clean and prepare the data for constructing the network models of tuning parameters.

\subsection{Developing Network Models of Tuning Parameters from Elogs}

\begin{figure}
    \caption{A schematic of the processes used to develop network models from the text data in the elog database}
    \includegraphics[width=1.0\textwidth]{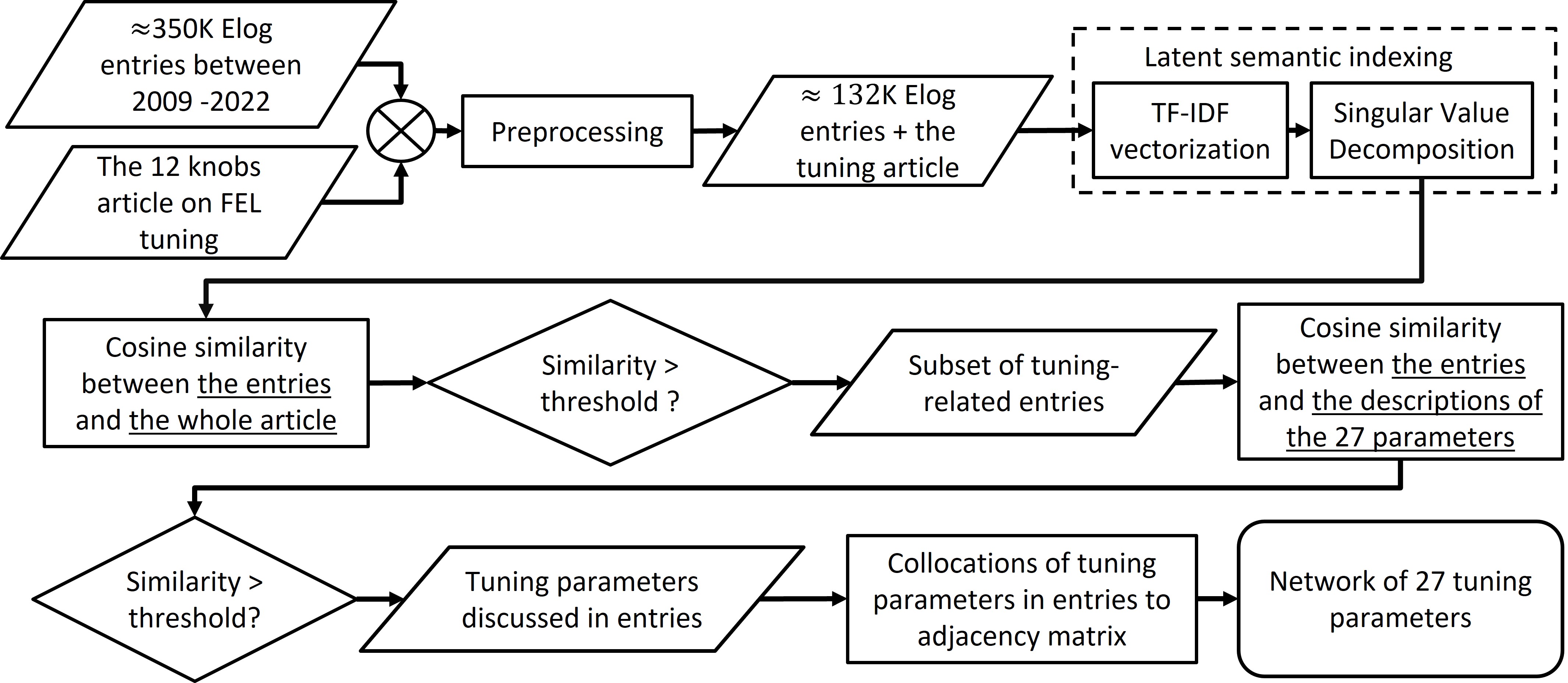}
    \label{fig:nlp_flowchart}
\end{figure}

Figure \ref{fig:nlp_flowchart} shows a schematic of our steps to develop network representation from elog data. At the pre-processing step, we remove (1) all machine logs (by author name), (2) any duplicate entries (by elog id), (3) the entries that contain less than 10 words with the title and the main text combined and (4) the entries tagged with labels from other accelerators. For each elog entry and the article, we remove all punctuations, URLs, email ids, stopwords, and numbers. After this step, we are left with about 132 thousand entries containing about 12 thousand unique words. We also use the same processes (when applicable) to prepare the FEL tuning training article.

Next, we represent the entries and the article as multi-dimensional vectors of all words or terms. We use Latent Semantic Indexing (LSI), a popular topic modeling method \parencite{manning2008introduction} that takes into account the context of each word in representing text documents as vectors. In addition, LSI also helps reduce the dimensionality of the data by mapping sets of words to topics. For the elogs, we reduce the dimensions of the entry vectors from 12000+ words to 100 topics (Please see Appendix \ref{app:lsi} for details). Then, we use the cosine similarity to find the similarities of the vectorized entries with the article on tuning to find the entries relevant to tuning. We use a threshold of 0.3 to classify each entry as relevant or irrelevant to the task. This process leaves us with approximately 2 thousand entries. Then, we once again use the tuning article – this time broken into the 27 parameters – to identify the topics of entries.

Finally, we use this information to find the collocations of the parameters in each entry and create an adjacency matrix based on the collocations. The adjacency matrices for the entries are added together for a set of entries (e.g., entries created by a group of operators). The resulting adjacency matrix for the set is used to construct undirected, weighted network models of tuning parameters as nodes. The edge weights between nodes are estimated from the collocations of the parameters in each entry.

\subsection{Network Analysis}

\begin{figure}[!ht]
    \begin{center}
    \caption{An example of a weighted network with 10 nodes and three communities (shown in different colors). The edge lengths represent the strengths of relationships between pairs of nodes. The solid edges denote in-community edges and the dashed edges represent out-of-community edges.}
    \includegraphics[width=0.5\textwidth]{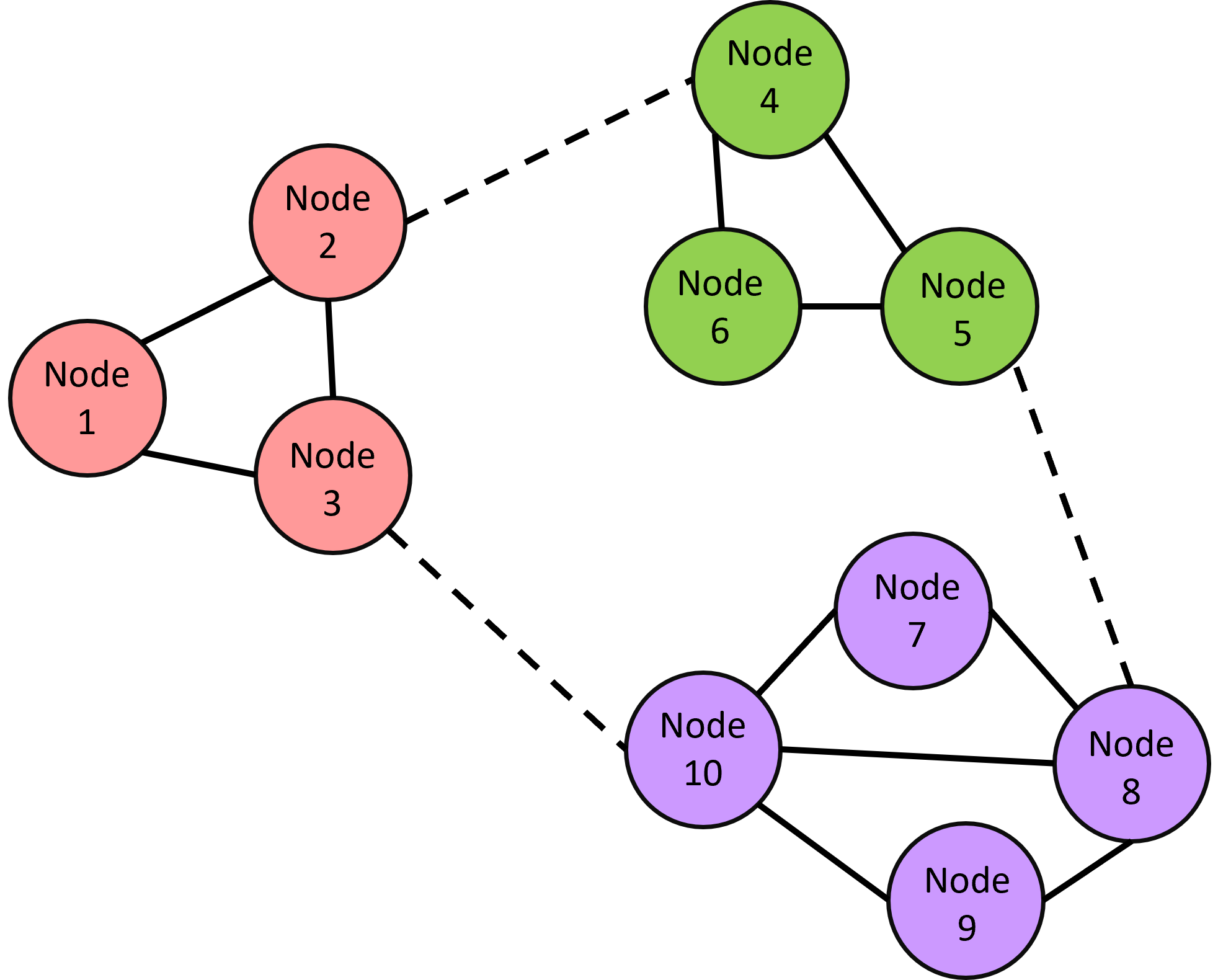}
    \label{fig:graph_illustration}
    \end{center}
\end{figure}

As discussed earlier, complex tasks can be analyzed at different levels of granularity. Figure \ref{fig:graph_illustration} shows an example network containing 10 nodes, 13 edges, and 3 clusters. Graph or network representations enable a host of network scientific measures and methods to study the changes at different network levels. Networks are constructed by defining the nodes and the edges between pairs of nodes. The nodes represent our variables of interest, and the edges represent different relations as attributes (e.g., weights, distances, correlation, transition probabilities). Thereafter, the networks can be investigated at different scales, ranging from microscopic views of the node or the edge importance in the network to more macroscopic views of the clusters of closely related nodes to whole networks.

In this work, we investigate the FEL tuning networks at four ascending levels of complexity: (1) Node level - individual subtasks, (2) Edge level - interconnections between subtasks, (3) Community level - groups of subtasks, and (4) Network level - the whole task.


\subsubsection{Node Level} 

Network science provides a variety of measures to examine a node's importance or role in a network from different perspectives \parencite[for a review, please see][]{siew2019cognitive}. For most of our analysis, we use the PageRank centrality measure (or simply PageRank) developed by \textcite{page1999pagerank}. It measures the relative importance of a node in the network based on the number of connections (or neighbors) and the importance of these neighbors themselves \parencite{page1999pagerank}. In other words, the PageRank of a node depends not only on the number of connections it has from its neighbors but also on the importance of these neighbors. 
PageRank is a specific variant of Eigenvector Centrality \parencite{graphbook}, that assigns importance to nodes based on its connectivity and the importance of its neighbors. The PageRank measure proved to be one of the most useful measures of node importance across various applications, such as from ranking webpages for Google searches to explaining human choices in a network of words in lexical retrieval tasks \parencite{siew2019cognitive, griffiths2007google}.

Other node-level measures we explored are (a) the degree centrality (a measure that assigns node importance based solely on the number of its connections, or its degree ) and (b) the clustering coefficient (which measures the probability of a node being in a cluster). However, we observed that these measures provide very similar distributions over the 27 tuning parameters as the PageRank. Therefore, these measures are not included in the main analysis, but are discussed with an example in Appendix \ref{app:detailed_networks}.

\subsubsection{Edge Level}

We use two measures of edge importance: (a) the edge weights and (b) the edge betweenness centrality. The edge weights express the strengths of connections between the nodes. Edge weights can be approximated in a number of ways \parencite[for reviews, please see]{siew2019cognitive, borsboom2021network}. In our case, the edge weights are estimated based on the explicit information of collocations of tuning parameters in text entries that we used in constructing the networks. This way, the edge weights represent the strengths of relations between node pairs.

The second measure, edge betweenness, represents the importance of an edge in connecting different parts of the network. It is calculated as the ratio of the number of shortest paths in the network that contain the edge and the total number of shortest paths in the network. We observed that the edge weights and the betweenness centrality portray opposing trends but similar rates of changes with expertise (please see Appendix \ref{app:detailed_networks} for a demonstration). For the brevity of this article, we include only the edge weights in the main text.

\subsubsection{Community Level}\label{subsec:communities}

Communities are local structures in networks, consisting of a group of nodes that have high edge density within the group and low density elsewhere. Community detection is an NP-hard problem, and therefore, finding optimal partitions is beyond the question with more than a few nodes. As an example, whereas there are 52 possible partitions for 5 nodes, there are 115975 possible partitions for 10 nodes. As the number of communities increases exponentially with the number of nodes, heuristic-based approaches are used for community detection \parencite{fortunato2010community, fortunato2016community}. 

In this work, we explored the presence of community structures using three heuristic-based algorithms: (1) the Girvan-Newman (G-N) algorithm, (2) the Louvain algorithm, and (3) Spectral Clustering. The strength of partitions of a network into communities is measured by the modularity metric developed by \textcite{newman2004analysis}, which provides a measure of the difference between the actual number of edges within a group of nodes, compared to the expected number of edges within the group in a random case. Modularity values range between [-1, 1], with modularity = 0 corresponding to the random case and positive modularity values indicating the presence of community structures in the network. Modularity values close to 1 have been rare in real networks, and values between 0.3-0.7 are considered to indicate strong partitions \parencite{newman2004finding, newman2004analysis}.


As a preview of our findings, the G-N algorithm failed to find strong partitions (with modularity $>$ 0.30). In contrast, the Louvain algorithm and spectral clustering showed consistent, strong partitions. In the main paper, we include the results of the Louvain algorithm which takes a hierarchical approach to optimizing the modularity metric itself. Please see Appendix \ref{app:detailed_networks} for details and demonstrations of the other two algorithms. 


\subsubsection{Structures Within Communities}
Notably, the communities represent sets of nodes that cluster together, but they do not provide information about the structure of the nodes within the communities or communities within the networks. Hierarchical clustering provides a suitable approach to examine the hierarchical structure of the networks. We use agglomerative hierarchical clustering based on linkage methods \parencite[]{hastie2009elements, james2013introduction}. For agglomerative clustering, we start with all individual nodes at the lowest level, and nodes based on pairwise distances as we go to higher levels. This combining process converges into communities and is the whole network at the highest level. 

The clustering results (i.e., the hierarchical structures) can be represented as dendrograms, and the distances at which nodes are combined indicate how far apart the nodes are in the network. The distances are usually calculated in an embedding space, taking either the maximum or the average distances between two groups of elements. Linkage refers to one such distance metric between two sets of elements. In this work, we adopted a ``complete" linkage method using the average distance between all pairs of points across two groups. Note that there are several other methods for hierarchical clustering we could have chosen as well, such as divisive clustering methods that start from the whole network and progressively divide into parts. Moreover, either the Louvian algorithm or spectral clustering may also be used in a divisive manner to achieve this goal. We chose the agglomerative method based on linkages due to its simplicity and popularity for hierarchical clustering.

\subsubsection{Network Level}
The network-level measures enable us to combine the information at different levels of the network into a single quantity. Many different measures to compare networks have been useful in different ways. For detailed reviews, please see \textcite{tantardini2019comparing, wills2020metrics}. We use two spectral distance measures -- (a) the adjacency spectral distance and (b) the Laplacian spectral distance -- due to their fast speed and reasonable accuracy to estimate the changes between networks.

\subsection{Measuring and Comparing Changes at Different Levels of Networks}


At the node and the edge levels, we get probability distributions of the measures over all possible options. To measure the changes at each of the two levels, we use two measures of distributional similarity – (a) Relative Entropy (RE) or Kullback-Leibler (KL) Divergence and (b) Overlapping Index (OI). RE is an information-theoretic measure of the difference between two distributions in terms of information contained in the distributions \parencite{cover2012elements}. The OI measures the amount of overlap in the areas contained between the distribution functions \parencite{pastore2019measuring}. Please see Appendix \ref{app:measures_of_change} for the equations.

At the community level, we find groups of nodes. We measure the changes using two similarity measures widely used in machine learning for testing clustering performance \parencite{rand1971objective, hubert1985comparing, vinh2009information}: (a) the Adjusted Rand Index (ARI) and (b) the Adjusted Mutual Information (AMI). The unadjusted versions of the measures range between 0 (all incorrect cluster assignments) and 1 (all correct cluster assignments). As the case of all incorrect assignments is unlikely and worse than random cluster assignments, the adjusted versions are corrected for chance and range between -0.5 (all incorrect assignments) to 1 (all correct), but now 0 corresponds to the expected clustering accuracy from random assignments.

At the network level, we use two measures of graph distance – specifically, the spectral distances based on (a)  the adjacency and (b) the graph Laplacian matrices of networks. The spectral distance between two graphs is calculated by the Euclidean or $L_2$ distance between the spectrums of eigenvalues of the matrices representing the networks. 

Please note that not all of these quantities are strictly metrics or measures, as they do not obey the triangular inequality. However, they are numerical indices well suited for our purpose of examining the changes in the measures over time. Another point to note is that some measures of change we use are measures of distance, whereas others are measures of similarity. Distance measures have a maximum of $\infty$ (zero similarity between distributions) and a minimum of 0 (identical distributions), whereas the similarity measures are scaled between 0 (no similarity) and 1 (identical distributions). To aid comparisons across levels, we convert all measures to distance measures in the main paper (Section \ref{sec:yearwise_changes}). The corresponding results through similarity metrics are included in Appendix \ref{app:year_by_year_with_similarity}. There are multiple ways to transform between the distance and similarity metrics. In this work, we use the Gaussian Kernel relationship ($Similarity = e^{-Distance^2} \Rightarrow Distance = \sqrt{-log_e(Similarity)}$) to convert between the similarity and the distance measures.

\section{Results}
\label{sec:results}

We begin by examining the changes at each network level (i.e., Whole Network, Community/Cluster, Interconnection/Edge Weight, and Subtask/Node) with years of experience. Thereafter, we examine the changes at each network level across three stages of operator expertise (i.e., Novice, Intermediate, and Expert).

\subsection{Changes in the Networks with Years of Experience}
\label{sec:yearwise_changes}

Before looking for network differences between our identified stages of expertise (i.e., Novice, Intermediate, or Expert), we look at the changes at a more continuous timescale of experience -- specifically, the changes with each half-year of experience. As the reference, we use the data from the first half-year and measure the changes across years relative to this reference period. At each network level, we use a measure of distance to investigate the changes.

\begin{figure}[!ht]
    \caption{Changes at different levels of the networks with experience. The first bin represents an operator's first six months and is used as a reference network for the distance metrics used to estimate the changes.}
    \includegraphics[width=1.0\textwidth]{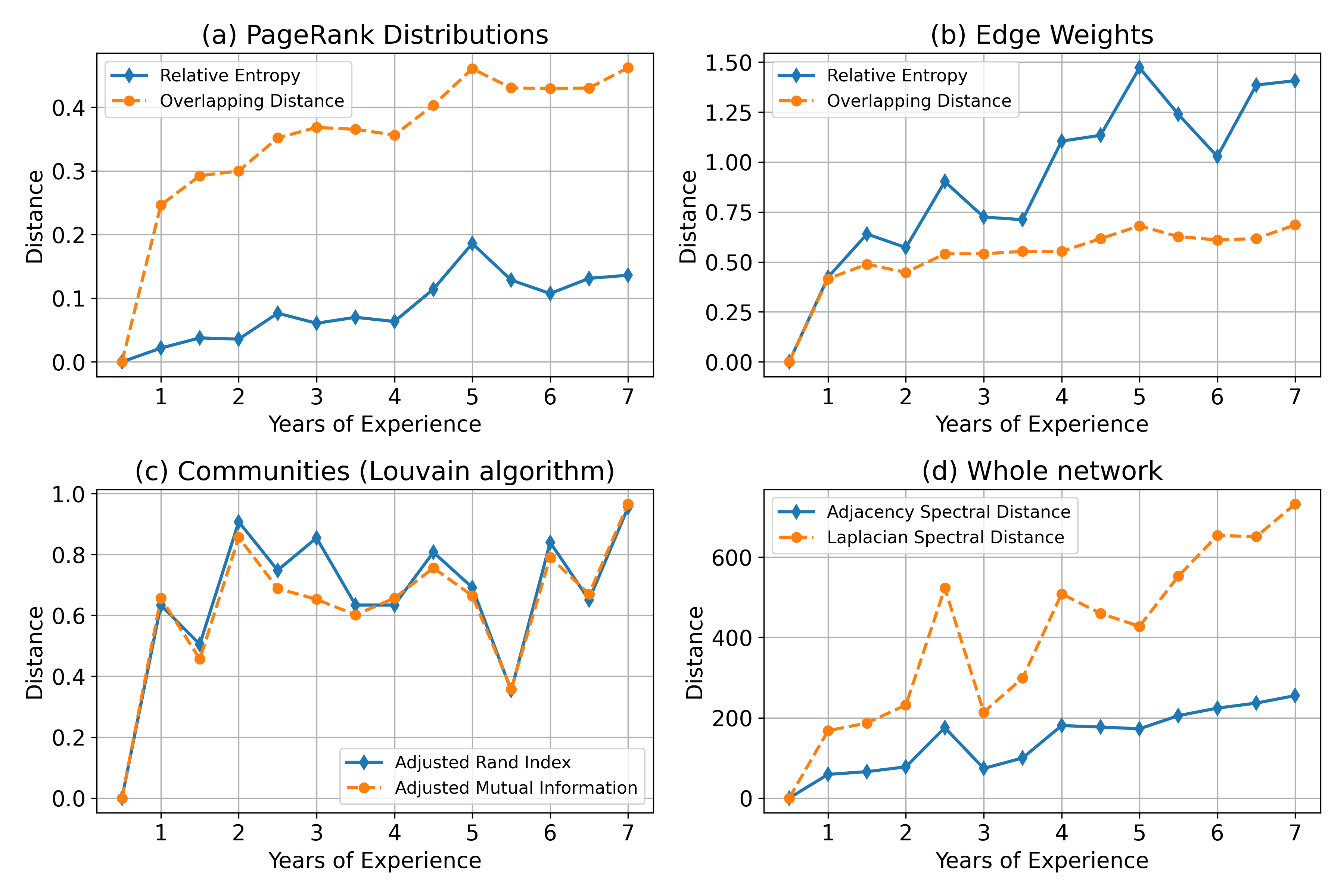}
    \label{fig:changes_over_years}
\end{figure}

The results are shown in Figure \ref{fig:changes_over_years}. The first two network levels (respectively, in Figures \ref{fig:changes_over_years}a \& b) are (a) the individual subtasks -- represented by PageRank distributions over the individual parameters, and (b) the interconnections among subtasks -- represented by edge weight distributions over all possible pairs of subtasks. As we obtain distributions for these network measures, we use two measures of distributional distances -- specifically (i) RE or KL Divergence (denoted in blue) and (ii) OI (denoted in orange).

For each of the two network levels, both RE and OI measures can be observed to reflect the same trend albeit at different scales. Using the first half-year as the reference network, the distances start from zero and increase as the dissimilarity between the network of interest and the reference network. Both the PageRank and the edge weight distributions change considerably with experience. At both the node and edge weight levels, the largest changes occur in the early years before the values eventually plateau with experience. Moreover, the peaks (e.g., at Years 2.5 and 5) and the troughs (at Years 2 and 6) of the two curves appear quite aligned with each other. For PageRank, the rate of change appears to be slightly lower than the rate of change in the edge weights (please note the different scales of the y-axis). This may reflect true changes in task selection are less pronounced than changes in how tasks are interconnected in the mental models of operators with expertise. There is also a possibility is that the different in growth rate between  subtasks and their interconnections is due to the two distributions having an unequal number of options in the distributions: 27 tuning parameters with 351 edges among them. Nevertheless, as the changes are largely continuous and uni-directional, the changes would accumulate over the years. Consequently, if there is a systematic change with expertise, they should be well captured in larger groups by binning across ranges of years, as we do in the next sections. 

To estimate the changes in the communities (Figure \ref{fig:changes_over_years}c), we use the ARI and the AMI values. Once again, we use the first period as the reference, and the curves start from 0. Both measures capture nearly identical changes. Unlike the PageRanks and the Edge weights, these changes seem quite discontinuous with experience. The largest increase in distance occurs in the first year. Barring this period, all distance values remain fairly consistent for all years. These results indicate that the margin of changes in the communities is smaller than at the node and the edge levels. As shown in the next sections, the communities remain similar even in our grouped investigation. In consultation with system experts, we identified that the three communities that arise within the networks across stages of expertise are consistent with groupings among the real-world tasks where each task falls into one of three main themes: 1) beam transport corrections, 2) beam energy and compression tuning, or 3) miscellaneous tasks. This alignment with our real-world problem helps validate the changes we see in the rest of our analysis. 


Finally, for the whole network (Figure \ref{fig:changes_over_years}d), we use the adjacency and the Laplacian spectral distances that measure the differences between two networks based on the adjacency and the Graph Laplacian matrices representing the whole networks. 
We observe that both spectral distances rise continuously with expertise, indicating that the whole networks change consistently as operators gain expertise. As the whole network encapsulates all of the network levels we have examined, these distances capture an aggregate of all the different ways the networks have changed over the years. A reflection of the aggregation can be observed in the peak at the 2.5 year mark and the trough at Year 3, resembling the PageRank and the edge weights (in Figures \ref{fig:changes_over_years}a \& b). On the other hand, there are also some notable differences. The peaks observed at Year 5 for the PageRank and the edge weights are not reflected in the changes of the whole network. Moreover, the rate of change in the networks appears much faster and more continuous than that of the PageRanks and edge weights. These differences in the rates of change suggest changes in the networks that are unaccounted for by the three lower levels we investigated. 

As a whole, these results indicate that there are measurable changes at all network levels that can provide valuable insights into the changes in FEL tuning strategy with expertise. The changes with experience in three of the four levels (excluding communities) are largely unidirectional. There are some similarities across the levels in terms of peaks and troughs. On the other hand, the rates of change seem different at different levels. We take a deeper dive into the changes at each level in the next sections by binning the operators into three groups: Novice (<1 year of experience), Intermediate (1-4 years), and Expert (>4 years).

\subsection{Whole Networks}
\begin{figure}[!ht]
    \caption{Networks of FEL tuning subtasks for three groups of operators. The node sizes represent the PageRank values for the nodes. The distances between nodes represent the edge weights. Communities were identified using the Louvain algorithm and verified using Spectral Clustering. }
    \includegraphics[width=1.0\textwidth]{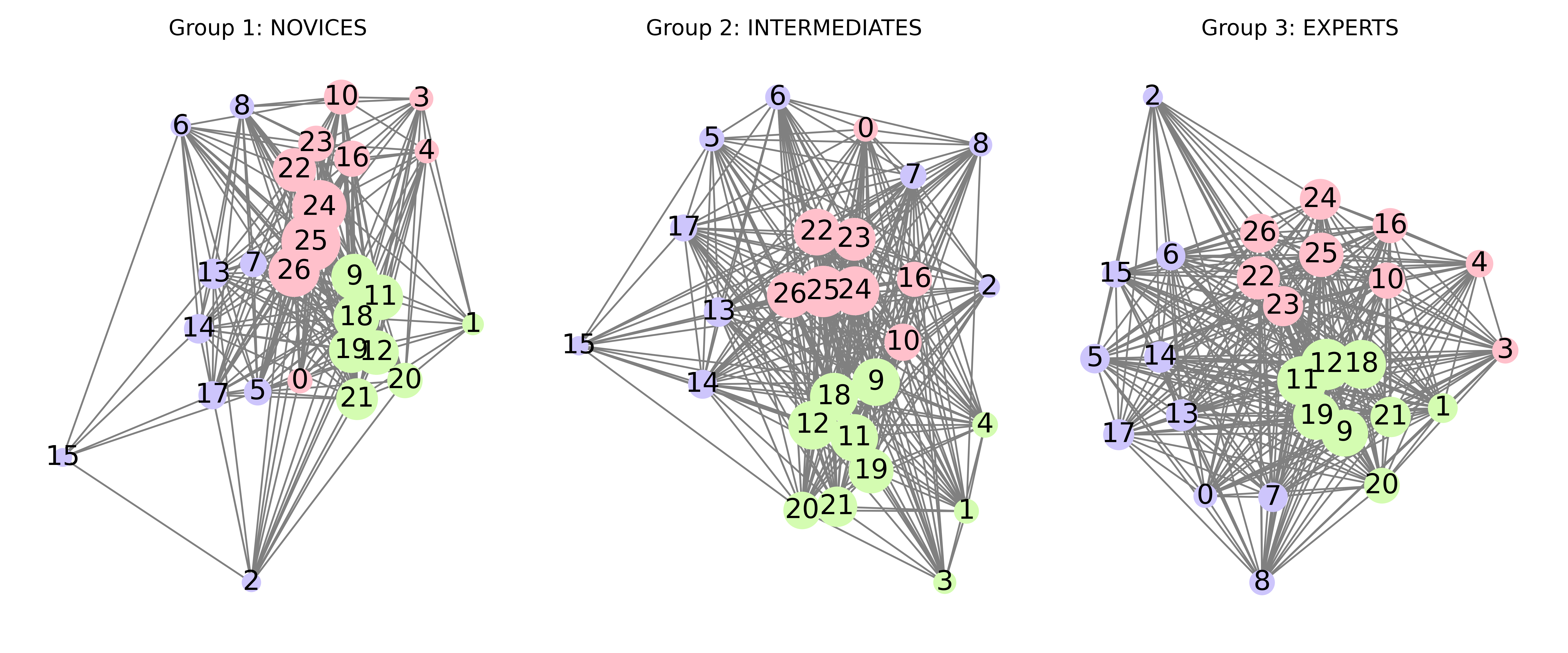}
    \label{fig:three_groups_whole_networks}
\end{figure}

Figure \ref{fig:three_groups_whole_networks} shows a bird's-eye view of the whole networks of FEL tuning subtasks for the three groups of operators. Each node represents one of the 27 task elements (i.e., tuning parameter to set), and their sizes vary proportionally to their PageRank. The networks are drawn using a spring layout where the edges act like springs. The spring constants are set proportional to the edge weights; therefore, nodes with high edge weights are pulled close together. The communities detected in each network are shown in different colors, and we see the same communities emerge for all three bins of expertise. The communities are well separated in the spring layout, providing some verification for the detected communities. We measure and examine these changes at different levels in more detail in the next sections, starting with the partitions of the networks into communities.

\subsection{Groups of Subtasks -- Communities and Hierarchies in the Networks}
\label{sec:communities_and_hierarchies}

We used the Louvain algorithm to detect the communities. This algorithm is a heuristic-based method that attempts to find the partition with the highest modularity. Note that for each network, modularity values of the partitions are above or close to 0.30 for all three bins of expertise. As discussed earlier, modularity indicates how unlikely the partitions are to emerge randomly. The range of 0.3-0.7 is recommended for a strong partition, as perfect modularity close to 1 is rarely found in real-world networks \parencite{newman2004analysis, newman2004finding}.  Notably, the three bins of operator expertise demonstrate remarkable similarities in how the subtasks are grouped into communities. The Louvain algorithm finds exactly three communities in the networks for all three groups. These communities are largely similar, with only one or two parameters being classified differently across groups (e.g., Parameter 0 for the experts and Parameters 3 \& 4 for the intermediates). Upon consulting with domain experts, we learned that the community denoted in green consists of parameters related to beam transport and steering, the purple community corresponds to parameters that affect beam energy and compression, while the pink set consists of all other parameters. In the next sections, we will denote the three communities respectively as (1) the beam transport (green) community, (2) beam energy/compression (purple) community, and (3) the miscellaneous (pink) community.

The similarities of communities demonstrated by the groups are quite striking, considering the large number of possible partitions ($\approx 5.45 \times 10^{20}$). These similarities strongly suggest that operators at all stages of expertise can recognize or categorize the parameters into similar groups. These results indicate that any differences in tuning performance with expertise are unlikely to stem from the improvements in the categorization of different parameters into communities. 

Though subtasks are consistently grouped into these three broad communities, there are differences in how subtasks converge into communities among the three stages of operator expertise. To more deeply examine how subtasks are organized into communities, we looked at the hierarchical structures of the networks using hierarchical clustering based on linkage methods that aim to group elements based on their distances in the network. By mapping the distances between subtasks, we can learn how community structures emerge among subtasks with expertise. 

\begin{figure}[!htb]
    \caption{Hierarchical clustering of networks of tuning performance and their changes with expertise. The height shows the distance between elements in the embedding space, representing the strength of interconnections between subtasks. The elements are clustered using this distance as a threshold. We keep raising the threshold distance and repeatedly check for elements with distances below this threshold and then group them together. The horizontal lines in the hierarchy represent the threshold distance at which the elements were grouped together. }
    \includegraphics[width=1.0\textwidth]{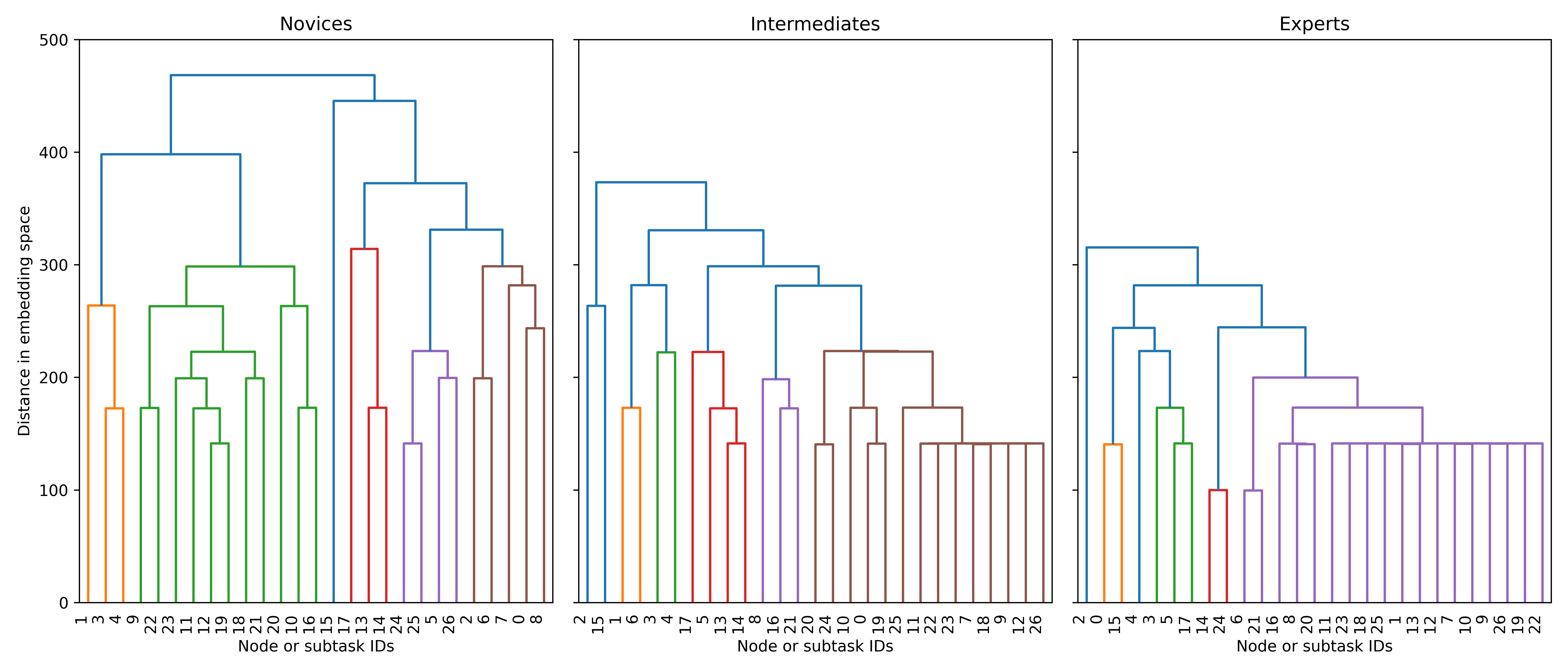}
    \label{fig:three_groups_hierarchy}
\end{figure}
The results from hierarchical clustering are shown as dendrograms in Figure \ref{fig:three_groups_hierarchy}. The height of the dendogram represents the distance between nodes in the network, which we can interpret as a strength of the interconnection between subtasks in the mental model of operators. We can see that the dendograms decrease in height with stages of expertise. This suggests that the subtasks are closer in distance and networks become more dense with expertise. Our results suggest that subtasks are more closely interconnected in expert operators' mental models and that subtasks converge into communities at lower distances. Second, we see that the networks become more well-structured, and the clusters become separated with expertise. The distance at which subtasks converge in the Novice dendogram is much higher and the elements are combined into many different sub-communities at low distances. In contrast, Intermediate and Expert operator dendograms show communities emerging at lower distances suggesting subtasks are more strongly organized in their models than those of novices. Finally, the changes in hierarchical structure from the Novice to the Intermediate states appear to be larger than those from the Intermediate to the Expert states reflecting the steep learning curve for Novice operators.

In the next section, we carry forward our knowledge of the communities and the hierarchies in the networks to investigate their edges, the interconnections between subtasks.

\subsection{Interconnections Between Subtasks -- Edge Weight Distributions}
\begin{figure}[!ht]
    \caption{The distributions of edge weights across all groups of expertise. It can be observed that the high edge weights correspond mainly to the in-community edges. The distributions also change slightly with expertise, showing the changes in the connections between task parameters with expertise.}
    \includegraphics[width=1.0\textwidth]{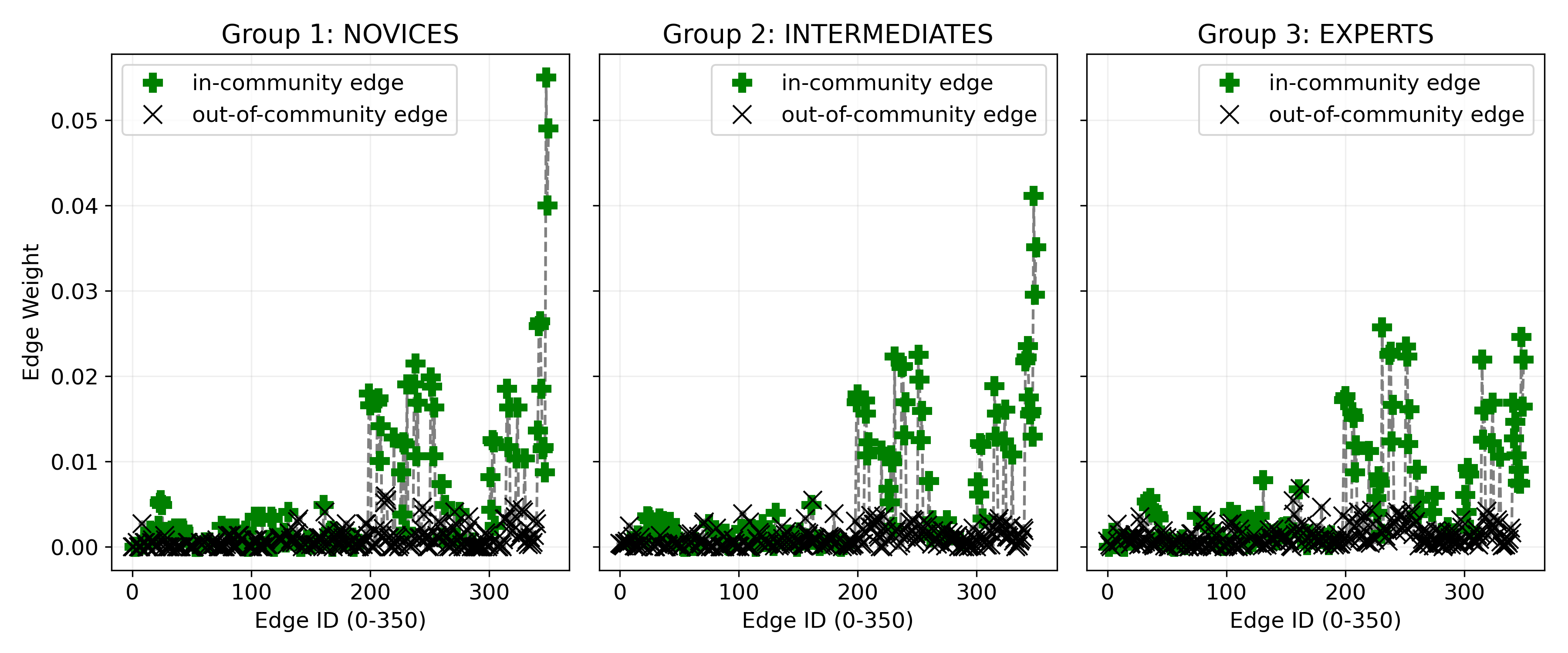}
    \label{fig:three_groups_edges}
\end{figure}

Figure \ref{fig:three_groups_edges} shows the distributions of the edge weights over all possible edges. The edges are presented in ascending order of connecting parameters or nodes (e.g., Edge 0 connects parameters 0 and 1, Edge 1 connects parameters 0 and 2, etc.) and the weight represents the distance between nodes.  We classify the edges according to whether they connect two nodes in the same community (in-community edges, denoted by green plus signs) or connect nodes that fall in different communities (out-of-community edges, denoted by black cross signs).

We see that the edges with high weights are predominantly in-community edges, and those with low weights are mostly out-of-community edges. To summarize these differences, we calculated the ratio of average weights for out-of-community vs in-community edges. We find that the ratios for the edge weights are 0.27, 0.31, and 0.43, respectively, for the Novices, Intermediates, and Experts groups. These results indicate that, although the three subtask communities are consistent among all stages of expertise, the Intermediate and, even more so, Expert operators have more densely interconnected networks as indicated by the larger out-of-network interconnection values. The change in these ratios with expertise is a consequence both of reduced in-network weights and increased out-of-network weights suggesting that Novices tend to stay more within communities when choosing the next subtasks than operators with more expertise. 




As for the distributional changes with expertise, a pronounced peak in the edge weights appears between subtasks 24 (Undulator Launch) and 25 (Undulator Orbit) for Novices, indicating these two tasks are strongly interconnected in their mental models. We can confirm from our hierarchical analysis that Novices connect these two tasks with the lowest distance. These two parameters are geographically close along the machine and involve the same system. Meanwhile, Experts make their strongest connection between subtasks 14 (BC1 compression) and 24 (Undulator Launch) which involve controlling devices over a mile apart in the accelerator. However, from an accelerator physics perspective, it is well understood that changing the beam compression at bunch compressor 1 (BC1) will affect the beam launch into the undulator, so it is natural to need to adjust subtask 24 after adjusting subtask 14. We can see evidence in our networks that the experts use accelerator physics domain knowledge to connect subtasks in their mental models. 

Apart from these differences, we see that the edges that have higher weights compared to the rest, remain quite similar across the four groups. For example, for each group, the edges labeled 200-250 and 300-350 seem to have the highest weights. We attribute the similarities at the edge level to the similarities of the communities. As the communities remain the same, the same sets of edges connect the elements within and outside of the communities. Generally, the distributions seem heavily influenced by the communities in the networks. Upon investigating the nodes connected by the edges (discussed in the next section) and consulting domain experts, we find the changes are most prominent in one community only – the one corresponding to miscellaneous subtasks. Taken together, we conclude that the importance of the individual edges in different aspects changes with expertise, but the changes appear to keep communities intact.

We also measured the changes in the distributions across the three groups. As we have only a few groups and the groups are of unequal sample sizes, we refrain from showing the changes in the distributions across these groups here. In section \ref{sec:yearwise_changes}, we will demonstrate the changes in the edge weight distributions with each year of experience.

\subsection{Individual Subtasks - Node PageRank Distributions}\label{sec:pageranks_results}
\begin{figure}
    \caption{PageRank distributions of the four groups. The subtasks are color-coded by the communities detected in earlier sections. }
    \includegraphics[width=1.0\textwidth]{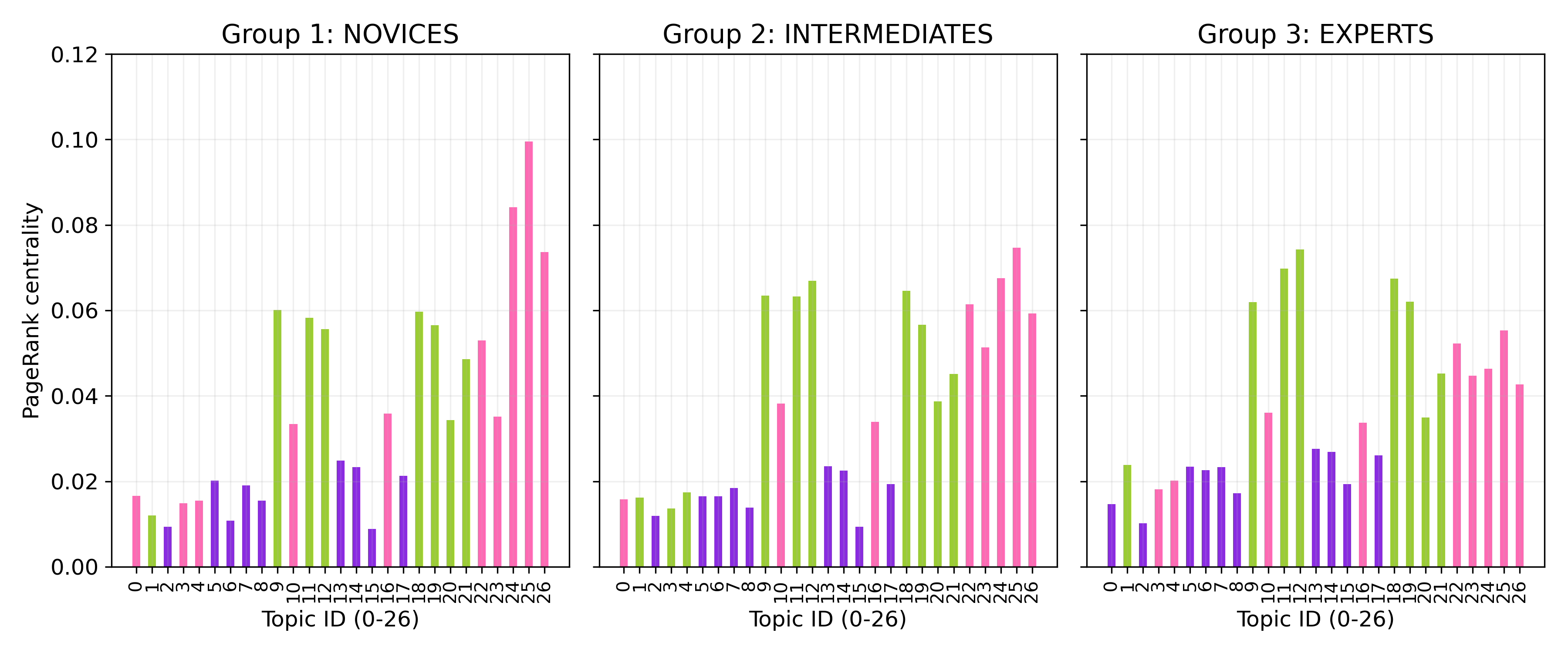}
    \label{fig:three_groups_pageranks}
\end{figure}

Now, we investigate the centrality of each individual node in the network using PageRank values, which we use to evaluate subtask importance. The PageRank distributions over the parameters are shown in Figure \ref{fig:three_groups_pageranks}. Each parameter is color-coded by the community they belong to, based on the Louvain algorithm we used for community detection (Figure \ref{fig:three_groups_whole_networks}). It is important to note that the PageRank value of a subtask represents a frequency as recorded in the elog and may be high either due to subtask importance or increased logging about the subtask. A task may appear overimportant in the Novice networks if the number of elog entries per execution of the subtask decreases with experience. 

As we see, the PageRank distributions show several interesting patterns relative to the communities in the network. For all groups, the beam transport (green) subtasks have consistently high PageRank values. This result makes sense since transporting the beam from the origin to the destination is a crucial step in generating photons. The PageRank values for the beam energy and compression (purple) subtasks seem to stay most consistent, growing only slights with expertise. However, there are large changes in the distributions for the Miscellaneous (pink) subtasks across stages of expertise. For Novices, this set contains subtasks with the highest PageRank values. 

Miscellaneous subtasks 24, 25, and 26 are all related to tuning the Undulator and show the most dramatic changes with expertise.  In addition to Undulator tuning subtasks, the changes in PageRank with expertise are most prominent for beam energy and compression subtasks 11 (Sector 21 Quadrupole Magnets) and 12 (BC1 Dispersion Quadrupole Magnets). The devices changed in these subtasks are located far upstream in the machine and the results from adjusting them are more dramatic and less predictable than other tuning subtasks. The PageRank values seem to indicate that operators gain confidence in choosing these subtasks with expertise. 

Finally, we observed in section \ref{sec:communities_and_hierarchies} that while the communities remain the same, the structures of the networks and the communities differ across groups. Therefore, it is possible that even if they divide the whole task into the same parts, operators of varying expertise differ in the underlying steps of task selection and sequencing-- resulting in the differences we see in the PageRank distributions.

\section{General Discussion}
\label{sec:discussions}

In this work, we adopted a network-based approach to model performance in the complex task of FEL tuning and investigated changes in these networks with expertise. Using a host of network scientific measures and methods, we examined the changes at four levels of the networks: (1) individual subtasks (nodes), the interconnections among subtasks ( edges), the groups of subtasks (communities), and the whole task (network). Our results reveal measurable changes with expertise across subtask, interconnection, and whole network levels. Across stages of expertise, subtasks are grouped consistently into three communities that reflect broad categories in the real-world complex task of FEL tuning, while the hierarchical structure of how substasks converge into communities reveals differences in how substasks are organized with operator expertise. 

In other words, the changes at the different levels of the networks and their structure suggest that expert strategies differ from that of novices. This change in how actions are connected enhances the ability of experts to perform the complex task of tuning beam brightness and represents a difference in how experts organize information that may translate to their strategy for other tasks in the control room. 

\subsection{Limitations of the Study}
We chose to reduce the complexity of the problem of operating the particle accelerator for our study by focusing only on the FEL tuning tasks out of the total tasks operators do in the control room. Since the selection and prioritization of FEL tuning tasks requires some strategy, we expected to see differences by expertise within this subset of tasks. However, the goal of maximizing photon beam brightness is not always the true objective of operators who are concerned with machine limitations, personnel safety, responding to failures, and other programmatic concerns. A holistic review of tasks would result in much larger networks with composed of many more subtasks. 

We used elog to non-invasively study expertise in the control room since real-time observations are difficult to coordinate in such a complex, high-stakes environment. The elog datastream is an imperfect collection of text entries describing control room events and excludes most machine state data and interpersonal interactions within the control room. Operators make a best effort to catalog tasks as they happen but the record is not a complete description of actions. Individual differences in syntax, detail, and event choice make the dataset highly variable. We also know that the actions of novices, including what to log, are supervised and guided by experts as they are being trained to operate independently. Expert input may skew the data and obscure some differences in strategy. 

We are limited in the machine learning methods we can use due to a relatively small dataset and a highly specific domain that make it difficult to train a model. Due to the highly specialized domain knowledge of this complex task, we worked with a very limited training data set for our NLP analysis of elog entries. The FEL tuning training articles written by experts that we used as reference collectively contain only  3500 words. In our similarity analysis between elog text and tuning reference articles, we may have inadvertently filtered out entries related to tuning that were missing reference keywords. 
Several of our mathematical methods are heuristic-based methods that do not guarantee best solutions. A main reason is the complexity of the problems that do not allow optimizing, ensuring that we cannot rule out the possibility that there may exist better solutions with alternative algorithms (e.g., for clustering) or metrics (such as in identifying the elog entries relevant to our topic of interest). Moreover, there are several subjective choices we had to make, such as the minimum number of words for in an entry and the entry-article similarity threshold to be included in our models. We mitigated algorithmic limitations by (1) using several alternative algorithms and metrics, and accepting results that are consistent across multiple alternatives, and (2) using bootstrapping methods to check the stability of the results from each alternative. In future studies, our methods need to be tested on a larger, multimodal dataset from more operational tasks and other complex tasks.

\subsection{Future Directions}
We hope that our work defines a useful methodology for studying human expertise in complex tasks. Within the control room setting, this analysis may offer ways to implement data-driven interventions in operator training or the control room environment, catalog control room tasks, and organize information like experts. 

In future studies, our network models of control room strategy can be fleshed out with more sources of information like machine state data and operator interactions with the control interface to help validate our findings and provide a more holistic view of the complex task. We can learn and add more subtasks to our networks and expand them to include control room tasks other than FEL tuning so that we can work with more of the elog entries that we filtered out for this study. Developing directional, multi-attribute networks can be immensely helpful, but extending our methods for such networks would require considerable development.

An enticing future direction is to observe how our measures evolve with time. We can analyze how time-intensive different subtasks are and use subtask durations to model operational tasks as sequences. There is much to learn about how the hardware and software environment, as well as administrative subtasks required for safe and efficient operations, impact the sequence and duration of subtasks in the control room. By incorporating more streams of data and exploring how subtasks are sequenced, we can not only learn something about the control room environment and operating procedures but also feedback on the training process to improve how training information is organized and delivered to trainee operators. 

Developing network models with different streams of information is itself an interesting problem. In this work, we used the collocations of parameters in the texts to estimate the adjacency matrices corresponding to the networks. We may estimate the networks based on different relationships between the parameters as nodes. For example, if performance data from actual FEL tuning sessions is available, we may directly use the operators’ action data to model performance as navigation in the tuning parameter network. In turn, comparing the networks developed based on different types of data and relationships can provide an even more detailed understanding of the learning processes of humans in complex tasks. 

Another future direction to pursue is building upon our methods. There are many different ways to represent graphs, measure their characteristics, and compare graphs to investigate changes in task selection and organization with expertise. In this work, we explored several methods that required expert validation and some feature engineering (e.g., estimating text similarity and extracting collocations). 
Another promising, more automated way to construct networks or graphs comes from graph-based machine learning (e.g., graph representation learning and graph neural networks). These methods have been helpful in capturing the evolution of networks \parencite{zhang2018link, zhang2022graph} and can help automate the network construction for the tasks we are investigating. 


 
\section{Summary and Conclusions}
\label{sec:conclusions}

We offer a new approach to study the changes of methods with expertise in complex tasks – by modeling task performance as networks of subtasks and measuring the changes at different levels of the networks. We test our network-based approach to investigate expertise in tuning particle accelerators. 

Our results reveal that the operators show continuous changes with expertise at the subtask, interconnection, community substructure, and whole-network levels. However, we also find that overarching subtask communities that emerge are remarkably similar among operators of all stages of expertise. These results suggest that the operators adopt a divide-and-conquer approach by dividing the whole tasks into parts of manageable complexity. We confirm with operations experts that these task communities reflect the real-world problem. Moreover, we also find that the changes across levels are not consistent with each other, rather occur at different rates and times. The changes at all levels except the communities indicate that although the divisions remain the same, the operators differ in the usage of the groups of subtasks. 

We propose this methodology as generalizable to different complex task settings and suggest that it is valuable to perform this type of analysis at all network levels to maximize understanding of the complex task. It may be that the preprocessing of data streams and choices of heuristics and measures for network design may vary greatly when using this methodology to study complex tasks in different domains. We highlight the importance of validating findings by checking that the networks built are representative of the complex task. This validation requires the precious effort of human domain experts. It is critical to identify key points in the network design or analysis that are validated by experts to ensure the networks are representative of the complex task studied and to intentionally select heuristics at each network layer that reveal useful information about changes in complex task strategy with expertise.

\section*{Code and Data Availability}
A Python implementation of all methods used can be found here: {\small \url{https://github.com/Roussel006/Expertise-in-Operating-Particle-Accelerators-through-Network-Models-of-Performance}}. The dataset can be found here: \url{https://osf.io/qmt2x/}.

\section*{Acknowledgments}
We would like to thank the accelerator operations team at SLAC National Laboratory who generously allowed us into the control room and shared their wealth of system expertise to made our work possible. Specifically, we would like to thank: the operators for giving us insight into their unique and incredible skill set and generating the elog data set; Peter Schuh, Johnny Warren, and Alex Saad for supporting and enabling this research and connecting us with the operators; Machine Specialists Benjamin Ripman and Janice Nelson for consulting on the design of this study with special thanks to Specialist Matt Gibbs who did all that and helped us navigate a myriad of networking and software development challenges. We would like to thank John Schmerge for being our upper-management advocate and extend our deepest gratitude to him and Mike Dunne for securing the additional funding we needed to complete this research. 

\printbibliography

\newpage
\appendix
\section{The 27 FEL Tuning Subtasks}
\label{app:tuning_parameters}

\begin{table}[!htb]
    \caption{List of FEL tuning parameters}
    \label{tab:tuning_params}
    \begin{tabular}{@{}l|rrr@{}}         \toprule
    ID 	& Parameters \\ \midrule
    0 	& LASER iris position \\
    1 	& Gun Solenoid Strength \\
    2 	& Schottky Phase \\
    3 	& CQ01 SQ01 \\
    4 	& XC01 YC01 \\
    5 	& 135 MeV bunch length \\
    6 	& LASER Pulse Stacker Delay \\
    7 	& LASER heater e- beam overlap (3D) \\
    8 	& LASER Heater Power \\
    9 	& Injector Matching Quads QA01 QA02 QE01-04 \\
    10 	& XCAV Launch Horizontal and Vertical \\
    11 	& S21 Matching Quads 21Q201 QM11-13 \\
    12 	& BC1 Dispersion Quads CQ11 CQ12 \\
    13 	& BC1 Horn Cutting \\
    14 	& BC1 Compression/Bunch Length \\
    15 	& 21-2 L1X Amplitude and Phase \\
    16 	& L2 Transverse Steering Launch \\
    17 	& BC2 Compression/Bunch Length \\
    18 	& BC2 Dispersion Quads CQ21 CQ22 \\
    19 	& Sector 26 Matching Quads \\
    20 	& DL2 Dispersion Quads \\
    21 	& LTU Matching Quads \\
    22 	& SXRSS Chicane Delay \\
    23 	& HXRSS Chicane Delay \\
    24 	& Undulator Launch \\
    25 	& Undulator Orbit \\
    26 	& Undulator Taper \\\bottomrule
    \end{tabular}
\end{table}

\newpage
\appendix
\section{Latent Semantic Indexing or Analysis (LSI/LSA)}
\label{app:lsi}

In this work, the entries and the article are represented as multi-dimensional vectors of all words or terms. The vector space representation allows several advantages, the most relevant of which is using the similarities between entries in the vector space for information retrieval or classification \parencite{manning2008introduction}. However, vector space representations based on just the counts or the frequencies of terms suffer from two common issues in NLP: (a) synonymy – many terms may have the same meaning, and (b) polysemy – one word may have different meanings. In this work, we use Latent Semantic Indexing (LSI), a popular topic modeling method that is well-equipped to deal with synonymy by taking the context of each term into account \parencite{manning2008introduction}. The problem of polysemy is mitigated in our case as the operators rarely use the same word to mean different entities. In addition, LSI allows reducing a document-word matrix (representing a whole database) from a high-dimensional space of words into a lower-dimensional space of topics by mapping sets of words to topics.

LSI consists of two steps: (a) Term Frequency-Inverse Document Frequency (TF-IDF) vectorization and (b) Singular Value Decomposition (SVD). In the TF-IDF representation for each document, the dimensions are the words, and the magnitudes are their TF-IDF values. TF of a term or word refers to the ratio between the number of times a word is used in a document and the total number of words in the document. IDF of a term is the logarithm of the ratio between the number of documents in the dataset (i.e., the elog entries and the tuning article) and the number of documents in which the term appears. The TF-IDF value for a term is the product of TF and IDF. Scaling by IDF helps to reduce the importance of terms that may occur very frequently (e.g., the term ``electron'' in FEL tuning) but convey little information about the specific document. 

The second step is performing SVD on the vectorized documents. The SVD process decomposes the original document-term matrix $M$ of dimensions $n$ terms $\times$ $m$ documents into a form of $M=USV^T$. The three factor matrices are (a) $U$, the term-topic matrix of dimensions $n\times k$, (b) $S$, a $k \times k$ diagonal matrix containing $k$ singular values representing the importance of the $k$ topics, and (c) $V^T$, a topic-document matrix of dimensions $k\times m$. The number of topics $k$ is the minimum of $(m, n)$ for the full case. To truncate, we set $k$ to be lower, essentially keeping only the highest $k$ singular values while zeroing out the rest. This way, SVD allows us to reduce the dimensionality of the data by mapping the words into a reduced number of topics. In our work, we truncate our original matrix of more than 12000 unique words using $k$=100 topics.

\newpage
\appendix
\section{Measuring Changes at Different Network Levels}
\label{app:measures_of_change}
\subsection*{Distributional Changes}
Suppose we have two distributions $p(x)$ and $q(x)$, and we aim to measure how different or similar they are. One of the most suitable and popular measures for such purposes is the \underline{Relative Entropy (RE)} or KL-Divergence \parencite{vedral2002role, cover2012elements}. RE measures the difference in the information contained in a target distribution, $p(x)$, that is not contained in a reference distribution, $q(x)$. The RE of $p(x)$ relative to $q(x)$ can be expressed as following \parencite{cover2012elements}: 

\begin{equation} \label{eqn:RE}
    RE (p(x)||q(x)) = \int_{X}\ p(x)\ \log_{a} \left(\frac{p(x)}{q(x)}\right)\ dx
\end{equation}

Some points to note about RE: (1) it is a relative and asymmetric metric; that is: $RE(p||q) \neq RE(q||p)$. A common technique (dating back to the original inventors of the metric) to make it symmetric is to average $RE(p||q)$ and $RE(q||p)$. We also used this technique in our investigations. (2) The minimum value of RE is 0, corresponding to $p(x)=q(x)$. The maximum RE is infinite. (3) The unit of RE depends on the base of the logarithm in Equation \ref{eqn:RE}. For 2-based logarithm, the unit is $bits$ of information, and for natural logarithm, the unit is $nats$ of information. In this work, we used natural logarithm to maintain consistency with the Gaussian Kernel used in transforming between the distance metrics and the similarity metrics.

Another metric we used is the \underline{Overlapping Index (OI)} developed by \textcite{pastore2019measuring}. The OI is dimensionless and directly measures the overlap between the areas contained by two distribution functions. The overlap between two distributions $p(x)$ and $q(x)$ can be expressed as following:

\begin{equation}\label{eqn:OI}
    OI(p(x), q(x)) = \int_{X} min(p(x), q(x)) dx = 1 - \frac{1}{2}\int_{X} |p(x) - q(x)| dx
\end{equation}

Whereas RE is a distance or dissimilarity metric, OI is a similarity metric. Minimum OI is 0 (corresponding to $RE\rightarrow\infty$) and 1 ($RE = 0$ and $p(x) = q(x)$). We use the Gaussian Kernel relationship to convert back and forth between the similarity and the distance metrics.

\begin{equation}\label{eqn:Gaussian Kernel}
    Similarity = e^{-Distance^2} \Rightarrow Distance = \sqrt{-log_e(Similarity)}
\end{equation}

\subsection*{Community or Cluster Changes}
In this work, we used two indices to measure cluster similarity: Adjusted Rand Index (ARI) and Adjusted Mutual Information (AMI). The details of these indices are beyond the scope of this work, but they can be found in \parencite{hubert1985comparing, vinh2009information, rand1971objective} among many other resources due to their popularity. The principles behind the indices are extremely simple to explain. Suppose we have a binary classification problem (Yes/No). For each element labeled positive (i.e., a ``Yes''), the label may be correct (True positive, TP) or incorrect (False Positive, FP). Similarly, each negative label may be correct (True negative, TN) or incorrect (False negative, TN). In this case, the \underline{Rand Index} (unadjusted) is simply the accuracy expressed by the proportion of correct labels (Equation \ref{eqn:RI}).

\begin{equation}\label{eqn:RI}
   RI(U, V) = \frac{Number\ of\ correct\ labels}{Number\ of\ total\ labels} = \frac{TP + TN}{TP+FP+FN+TN}
\end{equation}

Rand Index ranges from 0 (no correct labels) to 1 (all correct labels). For ARI, the index is adjusted so that (i) it ranges from -1 (no correct labels) to 1 (all correct labels), and (ii) an index of 0 corresponds to the expected number of correct labels when assigned randomly. This is accomplished as shown in Equation \ref{eqn:RI_to_ARI}, in which $E[.]$ denotes the expected value and $max(.)$ refers to the maximum value.

\begin{equation}\label{eqn:RI_to_ARI}
   ARI(U, V) = \frac{RI - E[RI]}{max(RI) - E[RI]}
\end{equation}

This approach can be easily extended beyond binary classification. Generally, for classification problems, our goal is to find a ``true" partition. A partition is a set of elements divided into subsets, clusters, or communities. For example, for a set $X=\{a,b,c,d,e\}$, two possible partitions among many are $p1 = \{ \{a,b\}, \{c,d\}, \{e\} \}$ and $p2 = \{ \{a\}, \{b,c,d\}, \{e\} \}$. When comparing two partitions, we estimate the accuracy based on the \textit{pairs of elements} in the clusters. Specifically, the pairs labeled correctly to be in the same cluster are the TP, the pairs labeled correctly to be in different clusters are the TN, and so on. To adjust, we use the expected number of correct labels in the random case. The full expression for \underline{ARI} between two partitions $P_1$ and $P_2$ of total $N$ elements can be expressed by Equation \ref{eqn:ARI} \parencite[][Chapter 3]{yang2016temporal} where $C_i$ and $C_j$ refer to the clusters in partitions $P_1$ and $P_2$ respectively. The other notations are: $N_i = N(C_i), N_j = N(C_j), N_{(i,j)} = N(C_i \cap C_j)$, and $N = N(C_i \cup C_j)$.

\begin{equation}\label{eqn:ARI}
   ARI(P_1, P_2) = \frac{ \sum_{(i,j)} {N_{(i,j)} \choose 2} \ \ \ \ \ \ \ \ \ \ \ - (\sum_{i} {N_{i} \choose 2} \times \sum_{j} {N_{j} \choose 2}) / {N \choose 2}}  { \frac{1}{2} (\sum_{i} {N_{i} \choose 2} + \sum_{j} {N_{j} \choose 2}) - (\sum_{i} {N_{i} \choose 2} \times \sum_{j} {N_{j} \choose 2}) / {N \choose 2} }
\end{equation}

\underline{Adjusted Mutual Information or AMI} between two partitions $P_1$ and $P_2$ can be expressed by Equation \ref{eqn:AMI} \parencite{vinh2009information}. $H(.)$ denotes information entropy, $MI(., .)$ denotes the mutual information between two partitions, $E[.]$ denotes the expected value, and $max(., .)$ denotes taking the maximum.

\begin{equation}\label{eqn:AMI}
    AMI(P_1, P_2) = \frac{MI(P_1, P_2) - E[MI(P_1, P_2)]}{max(H(P_1), H(P_2)) - E[MI(P_1, P_2)]}
\end{equation}





\newpage
\appendix
\section{Detailed Results from Network Analysis for Operator Groups}
\label{app:detailed_networks}

\begin{figure}[!htb]
    \caption{Network analysis for Group 1 consisting of Novice Operators}
    \includegraphics[width=1.0\textwidth]{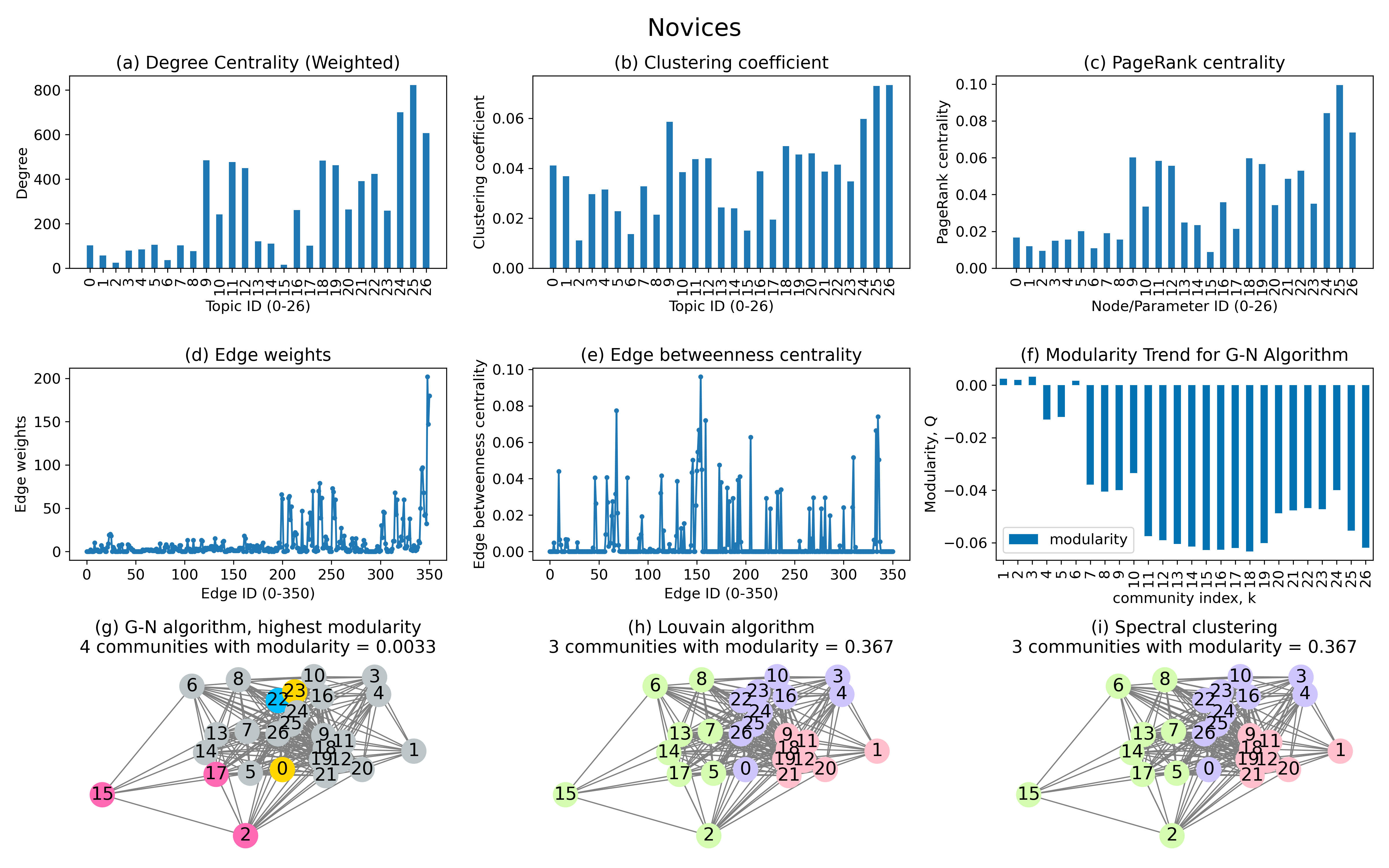}
    \label{fig:networks_for_novices}
\end{figure}

\begin{figure}[!htb]
    \caption{Network analysis for Group 2 consisting of Intermediate Operators}
    \includegraphics[width=1.0\textwidth]{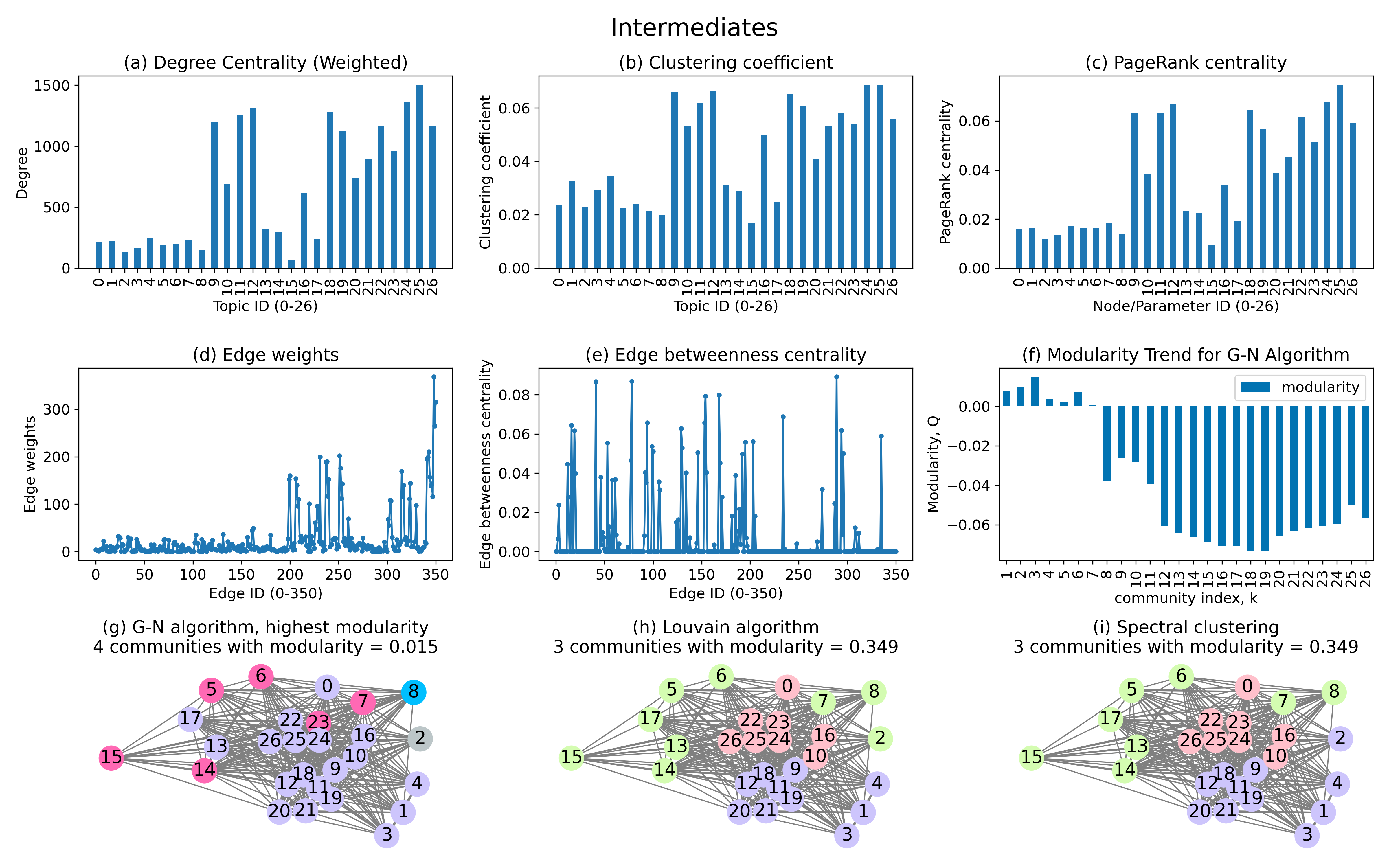}
    \label{fig:networks_for_intermediates}
\end{figure}

\begin{figure}[!htb]
    \caption{Network analysis for Group 3 consisting of Expert Operators}
    \includegraphics[width=1.0\textwidth]{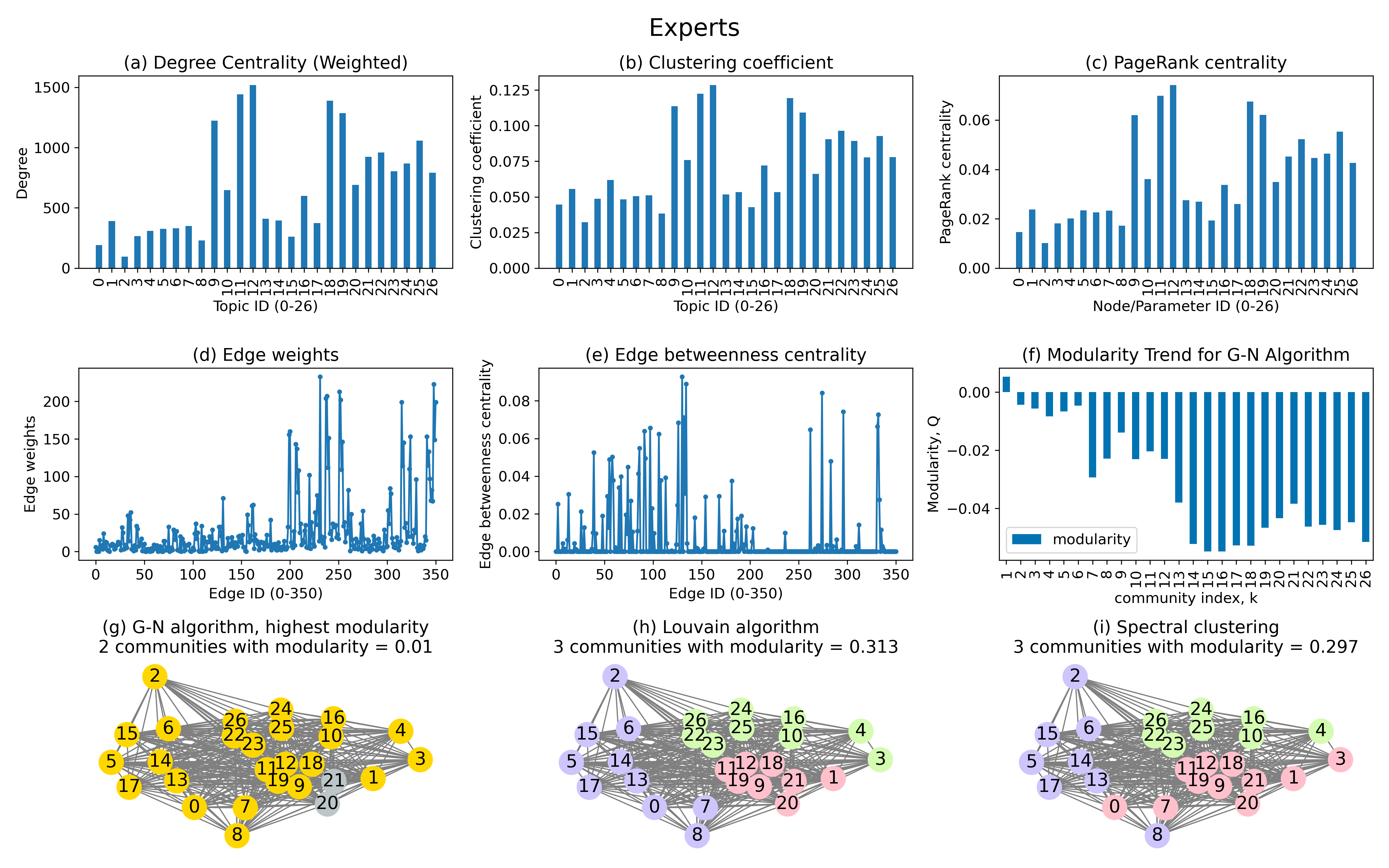}
    \label{fig:networks_for_experts}
\end{figure}

In this section, we present the detailed results from the network analysis of different groups of operators. In the main text, we focused on the changes with expertise at different levels of granularity using a specific set of measures. These measures at each level were selected after extensive testing since there is no singular “optimal” or best measure at any of the levels. Here, we demonstrate some of the measures we tried and how we selected the set we used in our analysis. In our first demonstration here (Figure \ref{fig:networks_for_all}), we include the tuning-related entries from all operators of all stages of expertise in developing the network. This example is discussed in detail. The detailed results for each of the three expertise groups discussed in the main text (i.e., Novices, Intermediates, and Experts) are included respectively in Figures \ref{fig:networks_for_novices}, \ref{fig:networks_for_intermediates}, and \ref{fig:networks_for_experts}.

\subsection*{Node Level}

\begin{figure}[!htb]
    \caption{Network analysis for all participants from all stages of experience. Panes (a)-(c) show the distributions of the node-level measures, respectively, the degree, the clustering coefficient, and the PageRank. As we see, the distributions are highly similar across measures. Panes (d) and (e) show the distributions of the edge level measures – edge weights and edge betweenness centrality – over all 351 edges (in the same order in Figures d and e). As we can see, there are some noticeable differences between the two distributions; for example, the edges with low weights appear to have high betweenness centrality. Panes (f)-(i) show the community level analysis: (f) shows the modularity trend of increasing partitions from the G-N algorithm; (g)-(i) show the communities identified by three different algorithms. Interestingly, while the G-N algorithm fails to find a strong partition of the network into communities (highest modularity  0), the Louvain and the spectral clustering algorithms both find a strong and very similar partition of the network (modularity > 0.3) into three communities.}
    \includegraphics[width=1.0\textwidth]{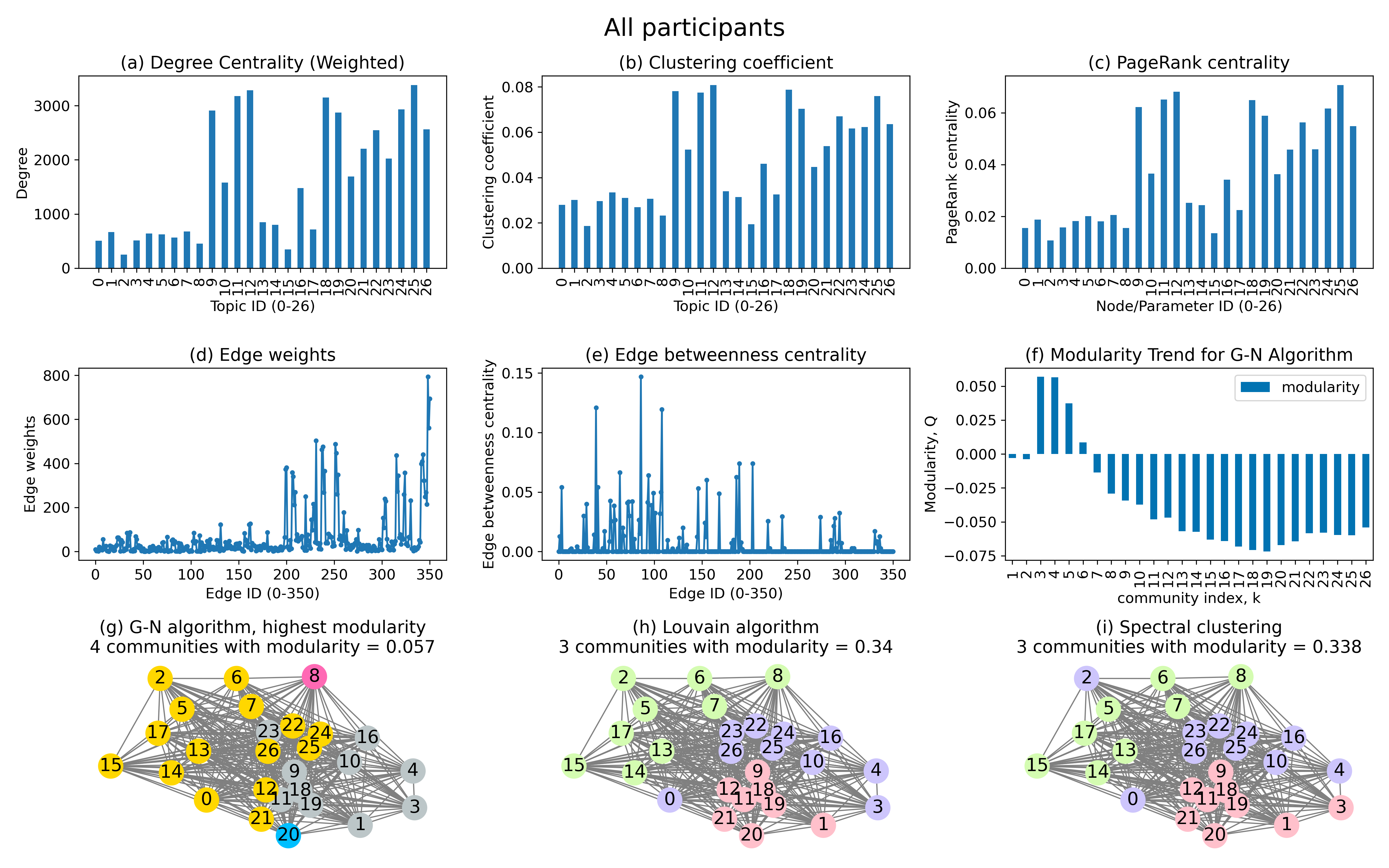}
    \label{fig:networks_for_all}
\end{figure}

We begin at the node level of the network – that is, the importance of individual parameters of FEL tuning. The three measures used at this level are the degree, the clustering coefficient, and the PageRank centrality of each node, shown respectively in Figures \ref{fig:networks_for_all} (a) – (c). The degree centrality (or simply degree) of a node captures the importance of a node by the amount of connections. The clustering coefficient of a node captures the probability or the tendency of a node and its neighbors to form (triangular) clusters. It is measured as the ratio of the actual number (unweighted networks) or the weights (weighted networks) of edges between the neighbors of a node and the maximum possible number or weight of connections between the neighbors. As networks with high clustering coefficients of the nodes are impervious to small changes (e.g., removal of an edge), the distribution of clustering coefficients provides an indicator of networks’ robustness to random changes.

The first point to note is that the three measures provide very similar distributions over the 27 nodes. For each measure, some parameters can be observed to be of lower importance than other nodes. For example, the values for parameters 9, 11, 12, and 18-26 appear to be higher than those of the rest of the parameters. These differences across parameters are the most prominent in the degree measure, followed by PageRank, and the differences are the least prominent in the clustering coefficient. A possible reason is that the PageRank of a node considers the information used in both the degree and the clustering coefficient of the node. The degree of a node is the total weight of the connected edges, whereas the clustering coefficient of a node represents the edge weights between its neighbors. The PageRank combines both sets of information by considering the importance of the neighbors of a node in estimating its importance.  As the distributions are fairly consistent across measures, we use only PageRank centrality as the node-level measure in the main analysis.

\subsection*{Edge Level}

At the edge level, we look at two measures: the weights and the betweenness centrality values of the edges. The results are shown in Figures \ref{fig:networks_for_all}(d)-(e). Please note that the edges are presented in the same ascending order using the IDs of node pairs connected by the edges. Unlike the similarity of the node-level measures, we see that the weights and the betweenness centrality measures demonstrate drastically different distributions over the edges. A general pattern appears that the edges with high weights are low on betweenness centrality, and the converse is also true. Moreover, the edges with high weights seem to flock together, and so do the edges with high betweenness centrality. The differences between the two measures can explain these differences. The edge weights represent the importance of connecting a pair of nodes, whereas the edge betweenness centrality values represent the role in connecting disconnected parts of the network. The contrasts in the distributions of weights and betweenness centrality indicate the presence of a strong community structure in the network, as the edges with high weights are likely to fall in the same community, and the edges with high betweenness are likely to be the connections between these edges.

\subsection*{Communities}

We confirm this possibility by investigating the communities in the network using three algorithms. The strength of partitions of the network into communities is represented by the modularity metric, which measures how densely connected a set of nodes is compared to the expected density of the nodes in a random case. As a brief reminder, a modularity value of 0 corresponds to the random case, and values greater than 0 indicate an escape from randomness and the presence of communities in the network. The recommended range for a strong partition is between 0.3 – 0.7. The first algorithm we use is the G-N algorithm, which creates communities by progressively removing edges in a descending order based on the betweenness centrality measure. Intuitively, the idea is that if island-like communities are bridged together by some edges, we will find the communities by removing these ``bridges". The distribution of modularity values for communities from the steps of the G-N algorithm is shown in Figure \ref{fig:networks_for_all}(f). As evident from all modularity values being close to or lower than 0, the algorithm fails to find a strong partition in the network. The communities from the best partition are shown in Figure \ref{fig:networks_for_all}(g) coded by different colors. While such low modularity values indicate a lack of community structures, it is also possible that the relatively simple principle of removing edges with high betweenness centrality contributes to the failure. A reason is that the G-N algorithm only considers a few possible partitions (specifically, the number of nodes minus 1) until it reaches a state where all nodes are singletons. However, there are close to $5.45×10^{20}$ possible partitions of the network. This possibility is also supported by the high number of edges with high betweenness centrality we observed in Figure \ref{fig:networks_for_all}(e), indicating that the communities in the network may be connected by several edges instead of one, which would have been ideal for the G-N algorithm.

The second algorithm we used is the Louvain algorithm (Figure \ref{fig:networks_for_all}h), another heuristic approach that aims to optimize the modularity metric hierarchically (Please see Section \ref{subsec:communities} for more details). As we see, the Louvain algorithm finds quite a strong partition with a modularity of 0.34 that divides the network into three communities. 

Finally, the third algorithm we used is spectral clustering (Figure \ref{fig:networks_for_all}i). The spectral clustering algorithm uses spectral decomposition of the original network and then uses closeness in the spectral domain to identify groups of nodes that cluster together. For binary classification, Fiedler values corresponding to the second eigenvalue in the spectrum are used. For more than two classes, using K-means clustering in the spectral domain is a common process, which is also the process we used in this work. As we see, spectral clustering yields the same partition as the Louvain algorithm. Three communities with a modularity value of 0.338. These results, along with our earlier investigations, confirm the presence of communities in the task network.
 
Another support for the success of the Louvain and the spectral clustering algorithms is present in the visualization of the network itself. The network plots in Figures \ref{fig:networks_for_all}(g) – (i) are drawn using a spring layout in which the edges act like springs with spring constants proportional to the edge weights, and the nodes with high edge weights are pulled close together. As we can see, the communities from both algorithms are well separated in the spring layout, lending an extra layer of support to the detected communities. As the Louvain algorithm and the spectral clustering algorithm produce very similar results, we use the results only from the Louvain algorithm when comparing the groups in the main paper. 



\newpage

\appendix
\section{Changes in the Networks through Similarity Metrics}
\label{app:year_by_year_with_similarity}
\begin{figure}[!h]
    \caption{Changes at different levels of the networks with years of experience with similarity measures. Please compare to Figure \ref{fig:changes_over_years}}
    \includegraphics[width=1.0\textwidth]{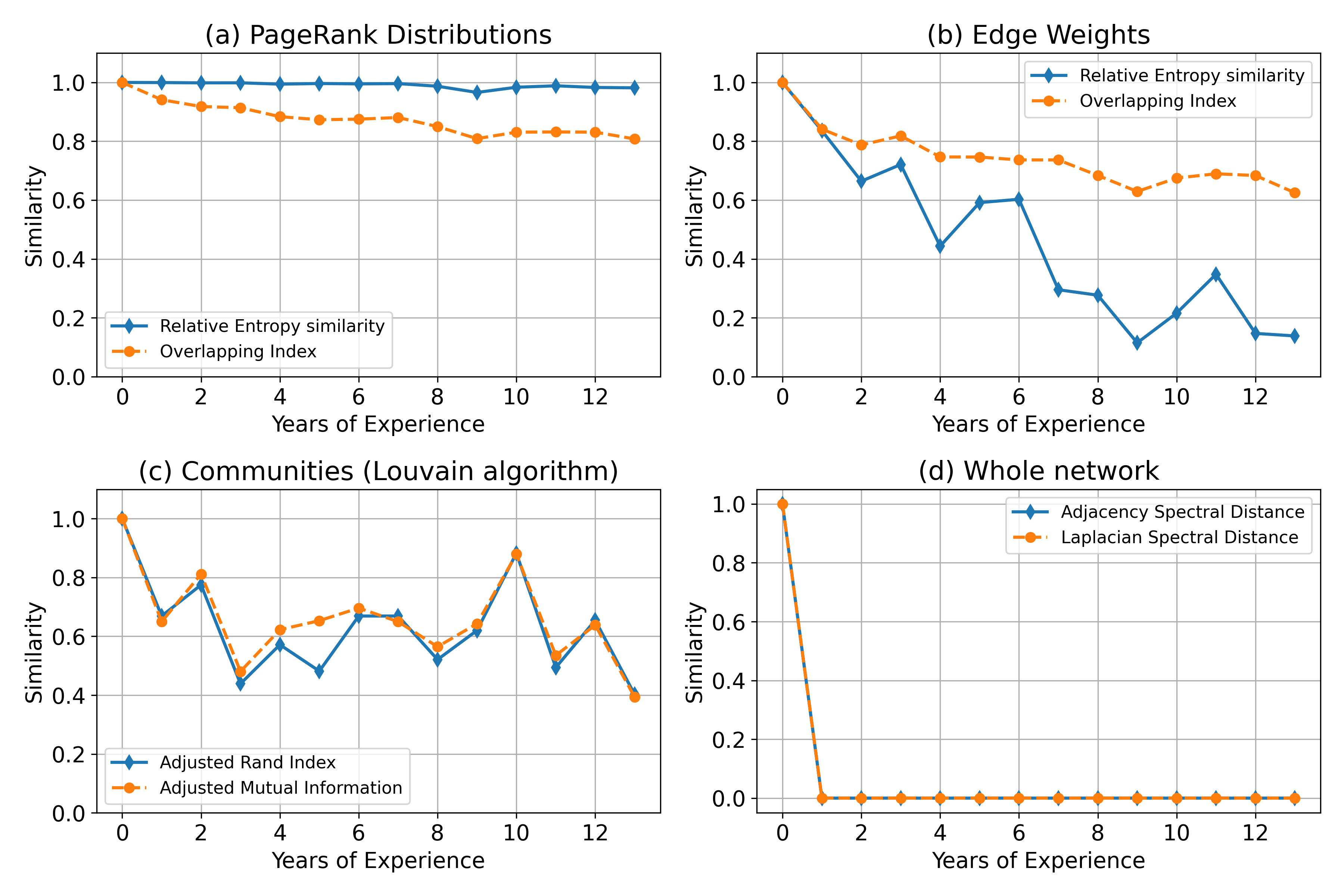}
    \label{fig:changes_over_years_through_similarity}
\end{figure}


\end{document}